\documentclass[aps,12pt,a4paper]{revtex4-2}
\usepackage{epsfig}
\usepackage{graphicx}
\usepackage{amsmath,amssymb,color}
\usepackage[english]{babel}
\usepackage{natbib}
\usepackage{graphicx}
\usepackage{float}
\parskip=\medskipamount
\usepackage{subfigure}
\usepackage{appendix}

\newcommand{\eq}[1]{(\ref{#1})}
\newcommand{\fig}[1]{Fig.~\ref{#1}}

\newcommand{\be}{\begin{equation}}
\newcommand{\ee}{\end{equation}}
\newcommand{\beq}{\begin{equation}}
\newcommand{\eeq}{\end{equation}}

\newcommand{\la}{\left<}
\newcommand{\ra}{\right>}

\begin{document}

\title{KPZ scaling from the Krylov space}

\author{Alexander Gorsky$^{1,4}$, Sergei Nechaev$^{2,4}$, and Alexander Valov$^3$}

\affiliation{$^1$Institute for Information Transmission Problems RAS, 127051 Moscow, Russia \\ 
$^2$LPTMS, CNRS -- Universit\'e Paris Saclay, 91405 Orsay Cedex, France \\ 
$^3$Racah Institute of Physics, Hebrew University of Jerusalem, Jerusalem, Israel \\
$^4$Laboratory of Complex Networks, Center for Neurophysics and Neuromorphic Technologies, Moscow, Russia}

\begin{abstract}

Recently, a superdiffusion exhibiting the Kardar-Parisi-Zhang (KPZ) scaling in late-time correlators and autocorrelators of certain interacting many-body systems, such as Heisenberg spin chains, has been reported. Inspired by these results, we explore the KPZ scaling in correlation functions using their realization in the Krylov operator basis. We focus on the Heisenberg time scale, which approximately corresponds to the ramp--plateau transition for the Krylov complexity in systems with a large but finite number degrees of freedom. Two frameworks are under consideration: i) the system with growing Lanczos coefficients and an artificial cut-off, and ii) the system with the finite Hilbert space. In both cases via numerical analysis, we observe the transition from Gaussian to KPZ-like scaling, or equivalently, the diffusion-superdiffusion transition at the critical  Euclidean time $t_{E}^*=c_{cr}K$, for the Krylov chain of finite length $K$, and $c_{cr}=O(1)$. In particular, we find a scaling $\sim K^{1/3}$ for fluctuations in the one-point correlation function and a dynamical scaling $\sim K^{-2/3}$ associated with the return probability (Loschmidt echo) corresponding to autocorrelators in physical space. In the first case, the transition is of the 3rd order and can be considered as an example of dynamical quantum phase transition (DQPT), while in the second, it is a crossover. For case ii), utilizing the relationship between the spectrum of tridiagonal matrices at the spectral edge and the spectrum of the stochastic Airy operator, we demonstrate analytically the origin of the KPZ scaling for the particular Krylov chain using the results of the probability theory. We argue that there is some outcome of our study for the double scaling limit of matrix models. For the case of topological gravity, the white noise  $O(\frac{1}{N})$ term is identified, which should be taken into account in the controversial issue of ensemble averaging in 2D/1D holography.

\end{abstract}

\maketitle

\section{Introduction}

The Kardar-Parisi-Zhang (KPZ) scaling is a highly universal phenomenon that has been observed in numerous physical problems and exhibits a rich mathematical structure. Recently, superdiffusion has been found in the late-time autocorrelators and correlators of several systems \cite{ljubotina2019kardar, scheie2021detection, bulchandani2021superdiffusion, ilievski2021superuniversality, de2020superdiffusion, PhysRevE.100.042116, krajnik2020kardar}. Numerical simulations suggest that the emergence of KPZ regime at late times is present both in integrable XXX spin chains and in their non-integrable deformations \cite{roy2023nonequilibrium, roy2023robustness}, indicating that this phenomenon is not ensured by the underlying integrability. It was hypothesized that the  KPZ scaling in correlators is linked to the formation of long-lived solitons \cite{de2020superdiffusion} or soft gauge modes \cite{bulchandani2020kardar}. A comprehensive review on this topic can be found in \cite{gopalakrishnan2023superdiffusion}, while the general review on the KPZ universality class is given in \cite{corwin2012kardar}.

This study delves into the emergence of KPZ scaling at late time from the perspective of Krylov space. The Krylov basis in the operator space can be considered for arbitrary seed operator encompassing both systems with a few or many degrees of freedom. The behavior of the Lanczos coefficients $a_n$, $b_n$ for any seed operator offers a quantitative description of some part of the system's Hilbert space. An important role of the Krylov basis for identifying the generic behavior of the interacting many-body systems was asserted in \cite{parker2019universal}. The time evolution of operators in the Krylov basis can be conceptualized as the motion of a quantum probe particle on chains with non-homogeneous hoppings, where each chain corresponds to a specific choice of an initial seed operator. Operator spreading at late times can be quantified through the Krylov complexity ${\cal K}(t)$, which provides the information about the typical distance along the chain, at a given time $t$, from the initial point of evolution. Different aspects of the operator evolution in the Krylov basis can be found in \cite{balasubramanian2022quantum, caputa2022geometry, rabinovici2022krylov, erdmenger2023universal, bhattacharjee2022krylov, bhattacharya2022operator, bhattacharyya2022towards, dymarsky2020quantum, jian2021complexity, caputa2021operator, hashimoto2023krylov, adhikari2023krylov,espanol2023assessing, dymarsky2021krylov}. The review of the application of Krylov basis for the different problems and the complete list of references can be found in \cite{nandy2024quantum}.

The behavior of the Krylov complexity at different timescales depends on whether the dimension of the Hilbert space of the model is finite or infinite. In our study we are interested in the finite-dimensional Hilbert spaces amounting to finite Krylov chains. The classification of the behavior of Krylov complexity at different timescales for finite Krylov chains has been done in \cite{barbon2019evolution, rabinovici2021operator}. It was argued that at small times up to scrambling time, fast growth occurs, and at the next stage, there is the linear growth up to the Heisenberg time $t_{hei} \sim K \sim e^{S}$ where $K$ is the length of the Krylov chain and $S$ is the entropy of the system. The large plateau starts at Heisenberg time when the Krylov complexity becomes saturated. Finally, after some Poincare recurrences, the complexity falls down at the very late double exponential time $t\sim \exp(\exp S))$. The time evolution of the Krylov complexity rhymes with the behavior of the Lanczos coefficients as the function of $n$. The scrambling regime and the linear growth of ${\cal K}(t)$ can be mapped into the growing dependence for $b_n$, while the plateau and the very late time decay of ${\cal K}(t)$ corresponds decreasing $b_n$ with $b_K=0$ at the end of the chain. In our study, we shall model these early and late time regimes by suggesting correspondingly that $b_n=(n/K)^{\alpha}$ and $b_n=(1-n/K)^{\alpha}$. The $\alpha$-dependence provides some intuition concerning the type of dynamical system.

Generically, the Lanczos coefficients are random variables, and it was argued in \cite{rabinovici2021operator} that it indeed happens if the physical system under consideration is integrable. The argument was based on the lack of interactions between the energy levels, which yields the Poissonian level statistics. The randomness of the Lanczos coefficients brings the problem of the Anderson localization with the non-diagonal disorder into the game. The relevance of the localization/delocalization on the Krylov space to the quantum chaos was conjectured in \cite{dymarsky2020quantum}, and it was shown in \cite{rabinovici2021operator} that the typical time of plateau saturation for the Krylov complexity is different for random and integrable systems. This argument has been checked for the XXZ spin chain in \cite{rabinovici2021operator} as well as for perturbed XXZ chains in \cite{rabinovici2022krylov}. Some aspects of the randomness of Lanczos coefficients and localization on the Krylov chain have been discussed in \cite{ballar2022krylov}.

The Krylov complexity has been recognized in the holographic approach: it quantifies the operator spreading in the radial coordinate discussed earlier in \cite{qi2019quantum} and can be identified via geometry \cite{caputa2022geometry}. The black hole represents the system with a finite number of degrees of freedom; hence, different regimes of complexity at different time scales are expected in this case as well. It was argued \cite{susskind2021complexity} that the scrambling regime and the linear growth have the clear interpretation when the probe particle falls toward horizon. However, the interpretation of the transition to the plateau and the plateau itself is a much more difficult task since, presumably, the quantum aspects of the black hole physics start to matter. In \cite{rabinovici2023bulk}, it was argued that there are indications that the wormholes start to play a role at the plateau and at the descending regime. However, the microscopic description of the transition from linear growth to plateau for the black hole complexity is still questionable.

The relation of the Krylov complexity with the bulk geometry in terms of the matrix model has been discussed in \cite{kar2022random} while the relation between complexity and the bulk volume for the operator growth problem framework was considered in \cite{jian2021complexity}. In the relatively simple situation of the double-scaled SYK model (DSSYK), one can identify the Krylov complexity for the finite-dimensional Hilbert space with the geodesic length in the wormhole in the thermofield double state. Moreover, the very Krylov basis has been identified with the basis of the fixed length geodesics in the two-sided $AdS_2$ \cite{rabinovici2023bulk}. It was argued from different perspectives that the non-perturbative wormholes are important at the Krylov complexity plateau or equivalently in the regime of descending Lanczos coefficients.

Another question concerning the possible relation of superdiffusion with 2D gravity is even more important. The 2D gravity interacting with matter was attacked by matrix models a long time ago, see \cite{di19952d} for a review. The large $N$ matrix model provides a kind of discretization of 2D geometry while the complimentary Kontsevich finite $N$ matrix model \cite{kontsevich1992intersection} yields the discretization of the moduli space. The $(2,q)$ minimal string is governed by the one-matrix model and it was found recently \cite{saad2019jt} that the Jackiw-Teitelboim (JT) gravity corresponds to the $(2,\infty)$ minimal string. Such identification of a 2D quantum gravity as the bulk theory with dual boundary matrix model \cite{saad2019jt,stanford2020jt} has induced the ongoing discussion whether the 2D gravity has to be considered as dual to a single boundary theory or to an ensemble of boundary models. There are many pro and contra arguments, however still there is no consensus at this issue.

Given such an interpretation of the matrix, the late-time asymptotics in 2D gravity are linked to the behavior at the spectral edge. In a 2D gravity context, the spectral edge behavior serves as the playground for the double scaling regime \cite{1990non,brezin1990exactly,douglas1990strings} where the summation over the genera is available. The double scaling limit in $(2,1)$ topological gravity is usually described via the spectrum of the deterministic Airy operator with the leading $\rho(\lambda)\sim \sqrt{\lambda}$ behavior at the edge. However, recently, this issue has been reconsidered in several aspects, aiming to take into account the non-perturbative physics responsible for the plateau regime. First, it was argued that the discreteness of the spectrum of the Airy operator has to be taken into account \cite{johnson2022microstate}, which uncovers the microstate structure via the Tracy-Widom statistics. Secondly, it was argued that subleading terms in the continuum spectral density carry the important information \cite{altland2024quantum}. It was conjectured in \cite{altland2024quantum} that the spectral edge is the point of the quantum phase transition (QPT) when the energy is the control parameter and the spectral density (which is non-analytic at the spectral edge) is the order parameter. 

In our study, we consider two cases of the finite Krylov chains:  i) all the Lanczos coefficients are growing and an artificial wall is introduced. In case ii), there are growing and descending Lanczos coefficient regimes along the complete Krylov chain. In both cases, we find the Gauss-KPZ transitions, which, however, are essentially different: the 3rd order transition in i) versus crossover in ii). We zoom the timescale in the vicinity of the Heisenberg time where both the Euclidean time $t_E$ and the Krylov chain length, $K$, tend to infinity simultaneously while maintaining the fixed quotient $t_{E}/K=c={\rm const}$ (for more details, refer to \cite{gorsky2018statistical}). Since at the Heisenberg time, it is expected that the correlations of fluctuations become non-linear, we search for the KPZ regime by varying the parameter $c$. We explore the late-time regime of correlators and autocorrelators by analyzing the Markov process associated with the probe particle dynamics on the finite Krylov chain. Remarkably, we have found the clear-cut transition to the KPZ regime at Heisenberg timescale $t_{E}^{*}=c_{cr}K$, where $c_{cr}=O(1)$. 

When considering numerically the one-point function, we select a position of a probe quantum particle at any arbitrary finite point on the Krylov chain and examine its evolution using the Liouvillian in the Krylov basis near the Heisenberg time for a prescribed inhomogeneity. In the late-time regime, above the critical time for $c>c_{cr}$, the terminal points of the particle's ``world lines'' are localized in a specific region of the Krylov chain, displaying KPZ-type fluctuations ($\sim K^{1/3}$), regardless the initial point. Conversely, at $c<c_{cr}$, the localization of final points on the Krylov chain depends on the initial points, and the fluctuations follow a Gaussian distribution. The very emergence of the KPZ superdiffusive regime is independent of the scaling exponent $\alpha$ in the Lanczos coefficient $b_n$, however the critical time value $t_{E}^{*}$ is $\alpha$-dependent, i.e., $c_{cr}=c_{cr}(\alpha)$. The emergence of the KPZ scaling for all $\alpha$ illustrates that the observed phenomena are not contingent on whether the system is integrable or not. This supports earlier observations in \cite{roy2023nonequilibrium, roy2023robustness} that the KPZ regime persists even when integrability is broken. 

Next, we examine the return probability on the Krylov chain (the Loschmidt echo) near the Euclidean  Heisenberg time. We observe that the return probability, which is the autocorrelator of the seed operator in the physical system, behaves as $\sim K^{-2/3}$ at $t>t_{E}^{*}$ in agreement with the dynamical scaling exponent in the KPZ regime. This change of behavior occurs at critical time $t_{E}^{*}=c_{cr}K$. Similar results are obtained at the Heisenberg timescale when distinct initial and final points of the probe particle are fixed.

For the full finite Krylov chain the transition looks as the crossover, however in the case i) we identify the 3rd order phase transition at the Heisenberg Euclidean timescale. There are clear similarities with the Dynamic Quantum Phase Transition (DQPT) when some observable, usually the Loschmidt echo, becomes non-analytic at the critical Minkowski time. This notion has been introduced in \cite{PhysRevLett.110.135704}, where it was suggested that the logarithm of the Loschmidt echo at the out-of-equilibrium is an analog of a free energy for the system at equilibrium. This quantity is non-analytic at DQPT, usually it develops a cusp in a 1D system. The reviews on DQPT and the list of references can be found in \cite{heyl2018dynamical, zvyagin2016dynamical, marino2022dynamical}. The search for the proper identification and classification of DQPT is currently under active study. In particular, it was demonstrated \cite{flaschner2018observation,heyl2017dynamical} that the creation of the topological defects in the momentum space takes place in the DQPT.
On the other hand, it was suggested recently that a kind of a ``long string'' spread in a spatial direction is created at the DQPT \cite{bandyopadhyay2023late}. The entanglement approach for the classification of DQPT has been discussed in \cite{de2021entanglement}. The DQPT has been studied in \cite{perez2022dynamical} in the matrix model context, and it was argued that it can be considered as the 3rd-order phase transition. We shall comment on this analogy and demonstrate that, indeed, some observable gets discontinuity at the critical time $t_{E}^{*}$. 

In case of descending Lanzcos coefficients it is possible (in the simple case) to recognize the emergence of the KPZ scaling analytically using of the properties of the Krylov basis for the matrix $\beta$-ensemble. The identification of the Krylov basis for the generic potential in the matrix model has been elaborated in \cite{balasubramanian2023tridiagonalizing} via the reformulation of the random matrix from $\beta$-ensemble as the tridiagonal matrix with non-trivial distributions \cite{dumitriu2002matrix} of the matrix elements. The parameter $\beta$ measures the strength of the noise. Hence, interpreting the matrix elements of tridiagonal matrices as the Lanczos coefficients, we can interpolate between the random Lanczos coefficients for finite $\beta$ and the deterministic case at $\beta \rightarrow \infty$. It was also shown recently \cite{PhysRevE.105.054121, das2023absence, das2024robust} that all possible regimes: localized, extended ergodic, and non-extended ergodic occur for eigenmodes in the $\beta$-ensemble for different values of $\gamma$ if the following scaling for $\beta$ is imposed: $\beta = N^{-\gamma}$. The parameter $\beta$ also serves as the coupling constant in the Calogero-Moser system \cite{awata1995collective} and the very partition function of the Gaussian $\beta$-ensemble is closely related to the norm of the Calogero-Moser ground state. Moreover, $\beta$ fixes the central charge of the auxiliary conformal field theory \cite{cardy2003stochastic} and the  deterministic $\beta\rightarrow \infty$ limit corresponds to the semiclassical limit $c \rightarrow \infty$.  

The KPZ scaling at the late Euclidean time evolution on the descending Krylov chain for the Gaussian ensemble with $\beta=1,2$ can be rigorously derived. The key observation made in \cite{ramirez2011beta} utilizing the previous results from \cite{edelman2007random} claims that near the soft spectral edge of the Gaussian $\beta$-ensemble the individual matrix from the ensemble can be approximated by the stochastic Airy operator (SAO), where the parameter $\beta^{-1/2}$ quantifies the strength of the noise. On the other hand, at $\beta=2$ the SAO and Airy process at large $N$ are related to the Cole-Hopf solution to the KPZ equation with the wedge initial condition \cite{gorin2018kpz, gorin2018stochastic, borodin2016moments}. The related results for the generic finite and infinite $\beta$ can be found in \cite{tsai2021large,gorin2020universal}. Another representation of the same solution of KPZ equation involves the moments of tridiagonal matrices at large $N$ \cite{gorin2018kpz}. These rigorous results from the probability theory provide the firm ground for KPZ scaling search in transition amplitudes along the generic Krylov chains in the simplest example.

The paper is organized as follows. Section II recalls the main notions concerning the Krylov space and the Krylov complexity, while Section III outlines the generalities concerning the KPZ universality class. In Section IV, we provide analytic arguments for the origin of the KPZ-like scaling on finite Krylov chains. Section V  is devoted to numerical results, showing the appearance of the KPZ-like scaling for the late-time evolution of the probe particle. Section VI discusses the implications of our findings for 2d gravity and derives the subleading at large $N$ white noise term in the double scaling limit. In Discussion, we briefly comment on the outcome of our study for the late-time correlators in the spin chains, as well as the possible link of the phase transition or the growing Lanczos case and the Euclidean version of DQPT. The results and open questions are summarized in Conclusion. The Appendix is devoted to the dependence on the non-vanishing Lanczos coefficient $a_0$.

\section{Krylov basis in the operator Hilbert space. Lanczos algorithm and Krylov complexity }

Here we recall the notion of the Krylov basis and the Krylov complexity in the operator space and emphasize the key features of the dynamics at the Krylov chain for the systems with finite number  degrees of freedom. Taking the arbitrary quantum Hamiltonian systems and selecting the particular seed operator $O$, we focus at its evolution in the Heisenberg frameworks, $O(t) = e^{iHt}O(0) e^{-iHt}$, which can be formulated in terms of the Liouvillian operator ${\cal L} = [H,\dots]$. Then, one applies the Gram-Schmidt orthogonalization procedure for the operators $\left.{\cal L}^n|O \ra$ which in this context gets reduced to the Lanczos diagonalization algorithm.

The Lanczos algorithm for the set of operators $\left.{\cal L}^n|O \ra$ in the proper operator basis goes as follows. Consider the recursion relations providing the orthonormal states $|O_n\rangle$ starting with $O_0=\left.|O\ra$, $O_1= \left.b_1^{-1}{\cal L}|O\ra$:
\begin{equation}
a_n =\la O_n|{\cal L}|O_n \ra,\qquad b_n=\la K_n|K_n\ra^{1/2}, \quad \left.|K_{n+1}\ra=\left.({\cal L}-a_n)|O_n\ra-\left.b_n|O_{n-1}\ra
\end{equation}
The orthonormal sequence of operators $\left.|O_n\ra=\left.b_n^{-1}|K_n \ra$ is called the Krylov basis in the operator space. The Lanczos coefficients $a_n,b_n$ depend on the seed operator, $O$, and form the tridiagonal matrix of the Liouvillian 
\begin{equation}
L_{nm}= \la O_n|{\cal L}|O_m\ra= \left(\begin{array}{cccccc}
a_0 & b_1 & 0 & 0 & 0 & \dots \smallskip \\
b_1 & a_1 & b_2 & 0 & 0 &  \smallskip \\
0 & b_2 & a_2 & b_3 & 0 &  \smallskip \\
0 & 0 & b_3 & a_3 & b_4 &  \smallskip \\
0 & 0 & 0 & b_4 & 0 &  \smallskip \\
\vdots & & & & & \ddots
\end{array}
\right)
\label{b04}
\end{equation}

The dynamics in the Krylov chain for the selected operator can be considered as the hopping problem for a particle along the one-dimensional semi-infinite or finite chain \cite{parker2019universal}
\begin{equation}
\frac{\partial \phi_n}{\partial t}= -b_{n+1}\phi_{n+1} +a_n\phi_n +b_n\phi_{n-1}
\end{equation}
where $\phi_n(t)= i^{-n}\la O_n(0)|O(t) \ra$, $\phi_n(0)=\delta_{n,0}$. The length of the chain depends on the dimension of the operator space. The hopping problem is equivalent to the Lax $2\times 2$ representation for the open Toda chain \cite{dymarsky2020quantum}, hence in the Toda language we discuss the behavior of the Baker-Akhiezer function $\phi_n(t)$. 

The important object is the autocorrelation function of the operator $O$:
\begin{equation}
C_{O}(t)=\la O|e^{i{\cal L}t}|O \ra
\end{equation}
which  can  be expressed in terms of the moments 
\begin{equation}
\mu_{2n}=\la O|{\cal L}^{2n}|O \ra = \frac{d^{2n}}{dt^{2n}}C(t)|_{t=0}; \quad C(t)=\sum_{n}\frac{\mu_n t^n}{n!}
\end{equation}
The moments themselves can be obtained via the product of $b_k$ along the properly defined Dyck's path $D_k$
\begin{equation}
\mu_{2n}= \sum_{h_n\in D_n}\prod_k b_{h_k +h_{k+1}/2}
\end{equation}

The Lanczos coefficients $b_k$ provide the essential information concerning the properties of the system. It was argued in \cite{parker2019universal} and checked in many examples that the large-$n$ asymptotic at $b_n= \mathrm{const}$ for infinite Krylov chain corresponds to the free system; at $b_n=\sqrt{n}$ -- to the integrable dynamics; and at $b_n=\rho n + O(1)$ -- to the chaotic dynamics. The coefficient $\rho$  turns out to be the important characteristics of the system. One can introduce the Krylov complexity \cite{parker2019universal, rabinovici2022krylov, balasubramanian2022quantum} which measures the average level in the hopping problem at time $t$ 
\begin{equation}
{\cal K}(t) =\sum_n n|\phi_n|^2
\end{equation}
For infinite chaotic system at $t\rightarrow \infty$ it behaves as
\begin{equation}
{\cal K}(t) = e^{2\rho t} 
\end{equation}
This behavior has to be considered as the quantum counterpart of the Lyapunov exponent what allows for the clear-cut geometrical interpretation \cite{caputa2022geometry} of the Krylov complexity. 

\begin{figure}[ht]
\includegraphics[width=0.4\textwidth]{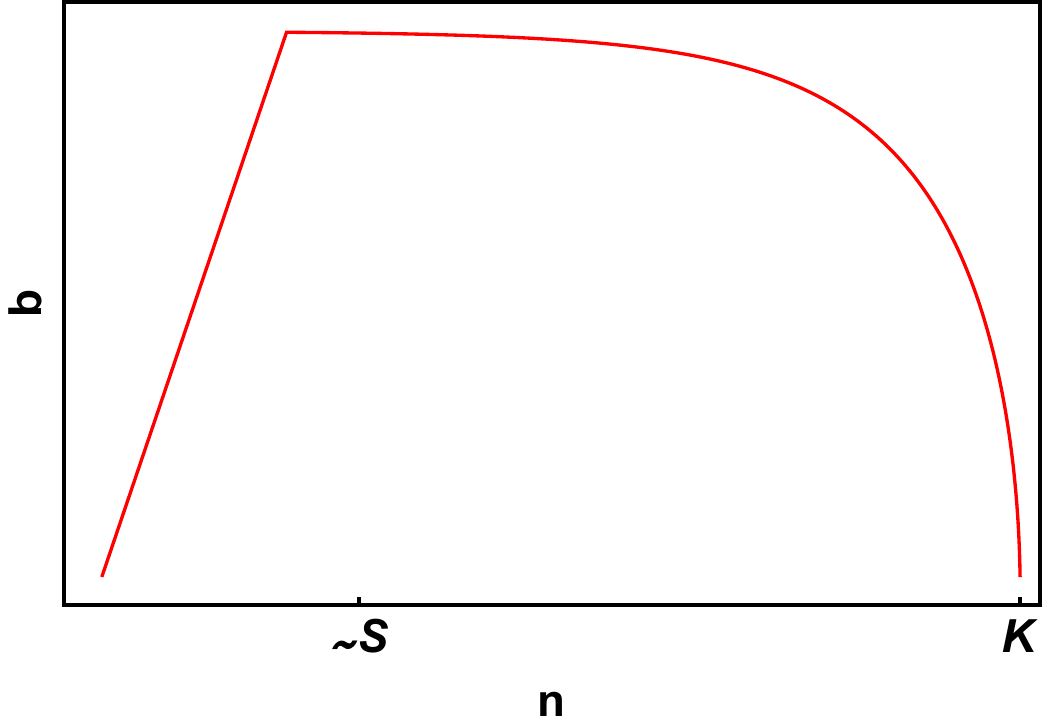}
\caption{Sketch of the general behavior of the Lanczos coefficients $b(n)$  for finite  system with $S$ degrees of freedom.}
\label{fig:000}
\end{figure}

In what follows we discuss the system with finite number of degrees of freedom $D$. In this case there are several important issues which play the key role. It has been argued in \cite{rabinovici2021operator} that the lengths of the Krylov chains $K$ obey the upper bound
\begin{equation}
K\leq D^2-D +1
\end{equation}
due to the properties of the spectral decomposition of Liouvillians. The finiteness of the Krylov chain length amounts to the universal behavior for the Lanczos coefficients at different time scales for the system with $D\sim e^{S}$:
\begin{itemize}
\item At times $0 \leq t \leq \log S$ the Krylov complexity grows in time, $b_n\sim \rho n$ for chaotic systems, and $b_n \sim \sqrt{n}$ for the integrable dynamics;
\item At times $t\geq log S$ the Krylov complexity grows linearly in time and $b_n$ is approximately constant;
\item At $t\sim e^{2S}\sim K$ the Krylov complexity is saturated and one has the descending regime in $b_n$ with $b_K=0$.
\end{itemize}
Below we zoom at the third timescale $t\sim K$ where the plateau in Krylov complexity begins. It is usually referred a bit loosely as the Heisenberg time, the term which we adopt here. We shall analyze the behavior of the system at the Heisenberg time scale. The behavior of $b_n$ for the system with finite-dimensional Hilbert space is sketched in \fig{fig:000}. Let us also emphasize that in this study we focus at evolution in the Euclidean time. The Euclidean time dynamics in the Krylov space rhymes with the dynamics in Minkowski time, however there are some subtle differences \cite{avdoshkin2020euclidean}.

\section{KPZ universality class}

Let us briefly recall some general facts regarding KPZ universality class which we exploit at length of the paper. Consider the growing interface in $(1+1)$ space-time separating the stable and unstable phases of some system. In a quite generic framework, the KPZ equation describes the interface dynamics for the height function $h(x,t)$
\begin{equation}
\partial_t h(x,t) = \nu \partial_{xx} h(x,t) -\lambda (\partial_x h(x,t))^2 + \xi(x,t)
\label{eq:burg}
\end{equation}
where $\nu$ corresponds to the surface tension $\lambda$ generates the non-linearity in the fluctuations
and $\xi(x,t)$ is the random noise in space and time. 

Two useful change of variables provide the additional flavors to the problem. Firstly, taking $u=\partial_x h$, \eq{eq:burg} can be converted into the form of stochastic Burgers equation
\begin{equation}
\partial_t u(x,t)+ \partial_x \left(-\lambda u^2 - \nu \partial_x u -\xi(x,t)\right)=0
\end{equation}
Secondly, the Cole-Hopf transform 
\be 
Z(x,t)=e^{\lambda h(x,t)}
\ee
brings the KPZ equation into the Stochastic Heat Equation (SHE):
\be 
\partial_t Z(x,t)=\tfrac{1}{2} \partial_{xx} Z(x,t) + \xi(x,t) Z(x,t)
\label{eq:she}
\ee
The function $Z(x,t)$ can be identified with the partition function of the directed polymer \cite{corwin2012kardar} in the Euclidean space-time plane. The first term in the rhs of \eq{eq:she} generates the Brownian motion and the second yields the interaction of the directed polymer with the random environment.

The solution to the KPZ equation depends on initial conditions. For flat initial condition $h(x,0)=0$ the late time asymptotics for one-point function reads as
\begin{equation}
h(0,t)\sim vt +(\Gamma t)^{1/3} \chi_{flat}
\end{equation}
where $\chi_{flat}=\chi_{GOE}$ with the probability distribution 
\begin{equation}
P(\chi_{GOE}\leq s)= \det \left(1- K_{\rm Ai}(s)\right) 
\end{equation}
and $K_{\rm Ai}(s)$ in the Fredholm determinant is the Airy kernel.

For the wedge initial condition $h(x,0)=|x|$ the probability distribution is introduced as follows
\begin{equation}
P\left(\frac{x^2}{2t}h(x,t) - \frac{t}{24} \geq -s\right)= F_t(s)
\end{equation}
and at large time it reduces to $F_{GUE}$
\begin{equation}
F_t(2^{-1/3}t^{1/3}s) \rightarrow F_{GUE}(s)
\end{equation}
The Tracy-Widom GUE distribution can be described as the Fredholm determinant as well
\begin{equation}
F_{GUE}(s)=\det (I -K_{\rm Ai}(s)) = \exp\left(- \int_{s}^{\infty} (x-s)^2 q^2(x)dx\right)
\end{equation}
where $q(x)$ obeys the Painleve II equation
\begin{equation}
q''(x)=\left(x+2q^2(x)\right)q(x)
\end{equation}
subject to the boundary condition $q(x) \sim {\rm Ai}(x)$ as $x \rightarrow \infty$.

The universality holds for the two-point correlators as well. For the stochastic Burgers equation the late time asymptotics of two-point function reads as
\begin{equation}
\la u(x,t)u(0,0) \ra\sim (\Gamma t)^{-2/3}f_{kpz}\left((\Gamma t)^{-2/3} x\right)
\end{equation}
where $f_{kpz}$ is fixed tabulated function which differs from the Gaussian form.
and $\Gamma = \sqrt{2}|\lambda|$.

\section{Krylov basis from the Gaussian $\beta$ ensemble perspective}

The late-time behavior $t\sim e^{2S}\sim K$ of the Krylov operator complexity in Euclidean time is governed by the descending regime of the Lanczos coefficients $b_n$. The instructive example for the Krylov chain with descending $b_n$ is provided by the  Gaussian $\beta$ ensemble.
It has been shown \cite{dumitriu2002matrix}, that random matrices from Gaussian (O)rthogonal/(U)nitary/(S)ymplectic ensemble ($\beta=1$, $2$ and $4$ respectively) can be represented in tridiagonal form: 
\be
H_\beta=\frac{1}{\sqrt{\beta K}}\left(\begin{array}{ccccc}
N(0,2) & \chi_{(K-1)\beta} & 0 & 0  & \dots  \bigskip \\
\chi_{(K-1)\beta} & N(0,2) & \chi_{(K-2)\beta} & 0  & \bigskip  \\
0 & \chi_{(K-2)\beta} & N(0,2) & \chi_{(K-3)\beta} & \bigskip \\
0 & 0 & \chi_{(K-3)\beta} & N(0,2)  & \bigskip \\
\vdots & & & &  \ddots
\end{array}
\right),
\label{b01}
\ee
where  $N(\mu,\sigma)$ and $\chi_{n}$ are the normal and the $\chi$-distributed random variables, respectively. Matrices $H_\beta$ have the same joint law of eigenvalues as matrices from  G(O/U/S)E: 
\begin{equation}
P(\{\lambda_i\})= \prod_{i<j}  \vert\lambda_i - \lambda_j\vert^{\beta}e^{- \beta \sum\lambda_i^2}
\label{jpdf}
\end{equation}
The ensemble of matrices $H_\beta$, referred to as Gaussian $\beta$ ensemble, provides a matrix representation of JPDF (\ref{jpdf}) for general $\beta$.

 The emerging tridiagonal matrix can be interpreted as the hopping Hamiltonian for the auxiliary particle propagating on the inhomogeneous discrete chain. It was found in \cite{balasubramanian2023tridiagonalizing} that the tridiagonalization procedure developed in \cite{dumitriu2002matrix} is nothing but the transition to the Krylov basis in the initial Hermitian matrix model. The Gaussian ensemble is not unique, and the transition to the Krylov basis can be developed for any measure in the matrix model, which yields the particular hopping $b_n$ on the Krylov chain. If the potential in the matrix model is even, then all $a_n=0$.

The tridiagonal maxtrix allows us to interpolate between the random Lanczos coefficients at finite $\beta$ and deterministic ones at $\beta=\infty$. Taking the limit $\beta \rightarrow \infty$, the matrix elements of the tridiagonal matrix become deterministic with the $b_n=\sqrt{1-n/K}, a_n=0$.

Let us recall the rigorous claim that the tridiagonal matrices $H_\beta$ from the Gaussian $\beta$ ensemble at arbitrary $\beta$ near the spectral edge are well approximated by the stochastic Airy operator $SAO_\beta$ \cite{ramirez2011beta}. The stochastic Airy operator is defined as follows
\begin{equation}
\mathcal{H}_\beta = -\frac{d^2}{dx^2}+x+\frac{2}{\sqrt{\beta}}B'(x)
\end{equation}
where $B(x)$ is the standard Brownian motion ( $B'(x)\equiv dB(x)/dx$  is the white noise). It has been shown that the largest rescaled eigenvalues of $\beta$ ensemble in the Krylov basis converge in distribution to the largest points of the Airy process  $SAO_{\beta}$ for any $k$ \cite{ramirez2011beta}
\begin{equation}
K^{2/3}\left(2-\lambda_{\beta,l}\right)_{l=1\ldots k} \underset{K\rightarrow \infty}{\to}( \Lambda_0(\beta) > \Lambda_1 (\beta)\dots> \Lambda_{k-1}(\beta))_{Airy} 
\end{equation}

Hence, in the limit $K\to\infty$ and $T>0$ the following holds:
\be 
\exp\left[-T K^{2/3}\left(2 I-H_\beta\right)\right]\underset{K\rightarrow \infty}{\to}\exp[-T \text{SAO}_\beta]
\ee
Simplifying, we get
\be 
\exp\left[T K^{2/3}H_\beta\right]\underset{K\rightarrow \infty}{\to}\exp[2T K^{2/3}-T \text{SAO}_\beta]
\ee
Hence the evolution operator $e^{-t H_\beta}$ at late-time behavior $t\sim e^{2S}\sim c K$, where the spectral edge makes the dominant contribution, converses to the  operator   
\begin{equation}
e^{c K H_\beta} \rightarrow \exp\left(2 c K-c K^{1/3}SAO_{\beta}\right)
\end{equation}

One more useful representation of $SAO_{\beta}$ found in \cite{gorin2018stochastic} involves the product of tridiagonal matrices  in $N\rightarrow \infty$ limit:
\be 
\lim\limits_{K\rightarrow\infty}\frac{1}{2}\left(\left(\frac{H_\beta}{2}\right)^{\left[T N^{2/3}\right]}+\left(\frac{H_\beta}{2}\right)^{\left[T N^{2/3}\right]-1}\right)= \exp\left(-\frac{T}{2}SAO_{\beta}\right)
\ee
where $[...]$ denotes the integer part of the variable.

Combining the relation between the asymptotic of the matrix product and $SAO_{\beta}$ from one hand, and the relation between $SAO_{\beta}$ and SHE-KPZ on the other hand, the authors of \cite{gorin2018kpz} have shown  that the Cole-Hopf solution to the Stochastic Heat Equation in the half space ($\beta=1$) or in the full space ($\beta=2$) for droplet initial condition provides:
\be
Z(0,2t^3)e^{\frac{t^3}{12}}=\lim_{K\rightarrow\infty}\frac{K}{\beta}\left[\left(\frac{H_\beta}{2}\right)^{2[tK^{2/3}]}+\left(\frac{H_\beta}{2}\right)^{2[tK^{2/3}]+1}\right]_{1,1}.
\label{g-s}
\ee
This relation is the matrix-element version of the operator correspondence discussed above.

A more general connection between solutions of KPZ  and the $SAO_{\beta}$ has been established in \cite{borodin2016moments} where it was obtained the exact relation between the Cole-Hopf solution to KPZ and the generating function of correlators for the Airy process at $\beta=2$:
\begin{equation}
E_{KPZ}\left[e^{-uZ(T,0)} e^{\frac{T}{24}}\right]=  E_{Airy}\left[\prod_{k=1}^{\infty} \frac{1}{1+u \exp (Ca_k)}\right] 
\label{eq:bor}
\end{equation}
with the identification of parameters $\frac{T}{2}=C^3$. This relation follows from the representations of the lhs and rhs in terms of the Fredholm determinants and the proper change of variables.

At the KPZ side \eq{eq:bor} reads
\begin{equation}
E_{KPZ}\left[e^{-uZ(T,0)} e^{\frac{T}{24}}\right] =1 +\sum_{k=1}^{\infty}\frac{(-1)^L}{L!}\int_{0}^{\infty} dx_1 \dots \int_{0}^{\infty} dx_L \det[K_u(x_i,x_j)]_{i,j=1}^{L}
\label{eq:bor1}
\end{equation}
where the kernel is as follows 
\begin{equation}
K_u(x,x')= \int _{-\infty}^{+\infty} \frac{dy}{1 + u^{-1} \exp((T/2)^{1/3} y)}{\rm Ai}(x-y){\rm Ai}(x'-y)
\label{eq:bor1a}
\end{equation}

At the Airy side the determinantal representation gives:
\begin{equation}
E_{Airy}\left[\prod_{k=1}^{\infty} \frac{1}{1+u \exp (Ca_k)}\right] = 1+ \sum_{k=1}^{\infty}\frac{(-1)^L}{L!}\int_{0}^{\infty} dy_1 \dots \int_{0}^{\infty} dy_L
\det[ K_{Ai}(y_i,y_j)]_{i,j=1}^{L}   
\label{eq:bor2}
\end{equation}
with the Airy kernel
\begin{equation}
\label{eq:bor2a}
K_{Ai}(x,y)= \int_{0}^{\infty} {\rm Ai}(x+a){\rm Ai}(y+a)
\end{equation}

Summarizing, equations \eq{eq:bor1}-\eq{eq:bor1a} on one side and \eq{eq:bor2}-\eq{eq:bor2a} on the other side, provide the explicit relation between the transition amplitude in the $\beta$-ensemble in the Krylov basis at the late time and the Cole-Hopf solutions to KPZ equation with the KPZ scaling behavior. Although this rigorous result is available for $\beta=2$ only, it provides the proper setup for numerical simulations at arbitrary $\beta$.

\section{Numerics}

In this Section, we elaborate on the behavior of the probe-hopping particle on finite Krylov chains. We shall first analyze numerically the descending part of the Krylov chain for the finite-dimensional Hilbert space represented by the tridiagonal matrix $H_\beta$ and find the Gauss-KPZ crossover both for deterministic ($\beta=\infty$) and random cases ($\beta<\infty$). Then, we introduce the Krylov chain with growing $b_n$ (typical for infinite-dimensional Hilbert space) and an artificial wall placed at level $K$ on the Krylov chain, which makes it finite. This system finds the Gauss-KPZ transition at the Heisenberg time of $t=c K$. We shall argue that the transition is of the 3rd order. Finally, we consider the full Krylov chain and argue numerically for different gluing of raising and descending parts, stating that there is no transition at the raising part, but there is a crossover at the Heisenberg time-scale for the descending part of the chain.

Let us explain briefly the simulation method. We approximate the evolution operator, ${\cal L}$, on a discrete Krylov chain in the following way: 
\be
{\cal L} = {\cal L}_{K\times K}= \left(\begin{array}{cccccc}
a_0 & b_1 & 0 & 0 & 0 & \dots \smallskip \\
b_1 & a_1 & b_2 & 0 & 0 & \smallskip  \\
0 & b_2 & a_2 & b_3 & 0 & \smallskip  \\
0 & 0 & b_3 & a_3 & b_4 & \smallskip  \\
0 & 0 & 0 & b_4 & a_4 & \smallskip  \\
\vdots & & & & & \ddots
\end{array} \right),
\ee
where $K$ is the length (number of sites) of the Krylov chain. 
In the continuous Euclidean time, $t$, the particle evolves on a Krylov chain via the Liouvillian operator $e^{t {\cal L}}$. To investigate the spatial characteristics of operator growth, we analyze the normalized transition amplitudes $P_t(i,j)$ between sites $j$ and $i$  ($i,j =\overline{1,K}$) of the Krylov chain
\be 
P_t(i,j)=\frac{1}{{\cal N}_j}[e^{t {\cal L}}]_{ij},
\ee
where ${\cal N}_j=\sum\limits_{i=1}^K [e^{t {\cal L}}]_{ij}$ is the normalization constant. The Krylov complexity for the hopping problem reads as
\begin{equation}
{\cal K}_i(t)=\sum_{j=1}^K  k P_t(i,j),
\end{equation}
where $i$ is a starting point of the particle evolution on the Krylov chain. The other important object is the correlation function, $C(j,t)$, which  can  be expressed in terms of the transition probabilities as follows:
\be
C(j,t)=P_t(1,j)
\ee

In what follows, we focus in our numerical simulations on the limit when Euclidean time is proportional to the size of the Krylov chain.

\subsection{Descending $b_n$}

In this section, we focus on studying how the probing particle evolves on Krylov chains with decreasing Lanczos coefficients $b_n$. We first consider the case where $b_n=\sqrt{1-n/K}$ and $a_n=0$, which represent the limit of $\beta=\infty$ of the tridiagonal matrix $H_\beta$. Additionally, we also explore the scenario where $b_n=(1-n/K)^{\alpha}$ to demonstrate that the appearance of the KPZ-like scaling is universal and a consequence of the behavior of the spectral density at the spectral edge. Hence, it does not depend on a particular matrix potential choice. Then, we show that the same scaling behavior arises for the case with fine $\beta$, i.e. when Lanczos coefficients are random.

\subsubsection{Deterministic Lanczos coefficients}

\begin{figure}[ht]
\centering
\begin{minipage}{0.4\textwidth}
  \centering
    \includegraphics[width=0.9\textwidth]{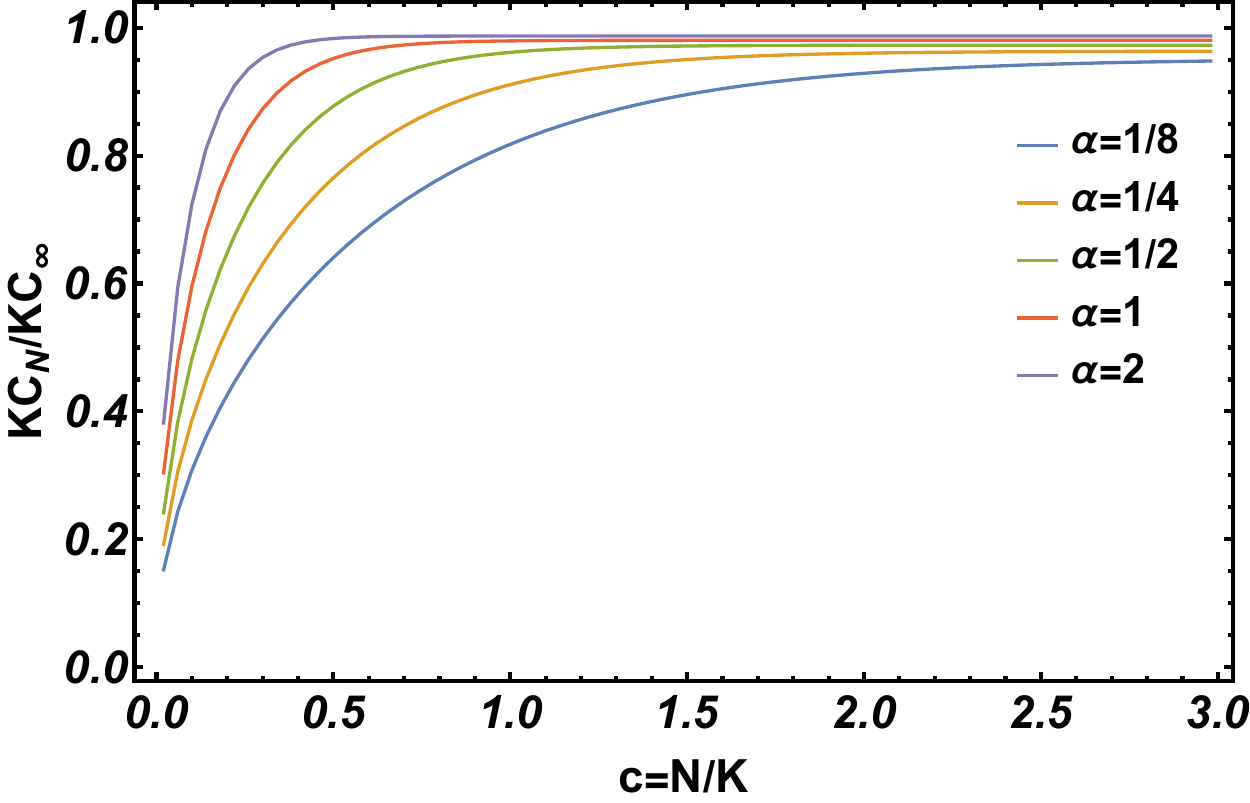}
\caption{Krylov complexity ${\cal K}(t)$ for a Krylov chain, where the hopping parameters are the Lanczos coefficients  $b_n=(1-n/K)^\alpha$. The plots for different $\alpha$ are presented.}
\label{fig:002a}
\end{minipage} \hspace{2cm}
\begin{minipage}{0.4\textwidth}
  \centering
  \includegraphics[width=0.9\textwidth]{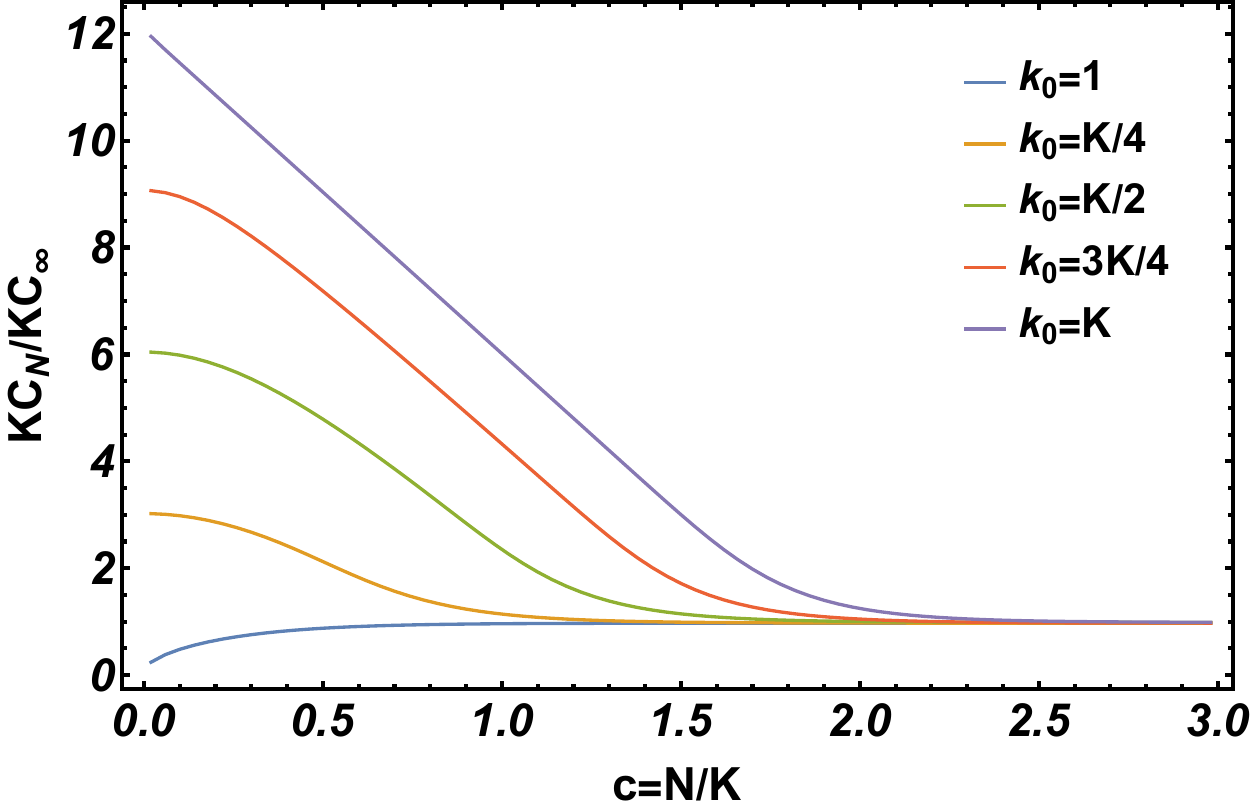}
\caption{Krylov complexity ${\cal K}(t)$ for a Krylov chain as a function of initial condition, $k_0$, where the hopping parameters are the Lanczos coefficients $b_n=(1-n/K)^{1/2}$.}
\label{fig:001}
\end{minipage}
\end{figure}

First, let us identify the KPZ scaling for the one-point function. That is, we start at $t=0$ at some point, $k_0$, on the Krylov chain with hopping coefficients $b_n\propto (K-n)^{\alpha}$ and perform the Feynman-like path integral over all paths of time $t$, assuming the relation $t=cK$. It turns out that at the Heisenberg timescale, the distribution of the ends of the trajectories depends on the value of $c$. At $c>c_{\text{cr}}$ all trajectories irrespectively on the initial point end up in the strip of width $K^{1/3}$ around the asymptotic value of Krylov complexity ${\cal K}_{as}$ (see \fig{fig:002a} and \fig{fig:001}). The mean value of  the ends of trajectories is parameterized by the asymptotic value of Krylov complexity
\begin{equation}
\label{strip}
{\cal K}_{as}\sim (K/\alpha)^{1/3}
\end{equation}
for $b_n\propto (1-n/K)^{\alpha}$. The width of the localization strip, which quantifies the fluctuations, follows the KPZ-like scaling of $K^{1/3}$.

At the qualitative level, the asymptotic value of Krylov complexity (\ref{strip}) can be derived using the approach developed in \cite{parker2019universal, yates2020dynamics}. Let us consider the continuum approximation of the hopping problem when the inhomogeneous hopping can be substituted with the Dirac equation for the probe spinless fermion with a varying mass. The mass dependence can be read from the effective potential $V(x)=(1-x/K)^{\alpha}$. The summation over trajectories we performed can be considered as the path integral derivation of the wave function in this Dirac equation. The solution to the Schrodinger equation at $c>c_{cr}$ yields $\langle x\rangle$ for this potential coincides exactly with (\ref{strip}) and the observed width of the strip around $\langle x\rangle$ is reproduced as well.

\begin{figure}[ht]
\includegraphics[width=0.9\textwidth]{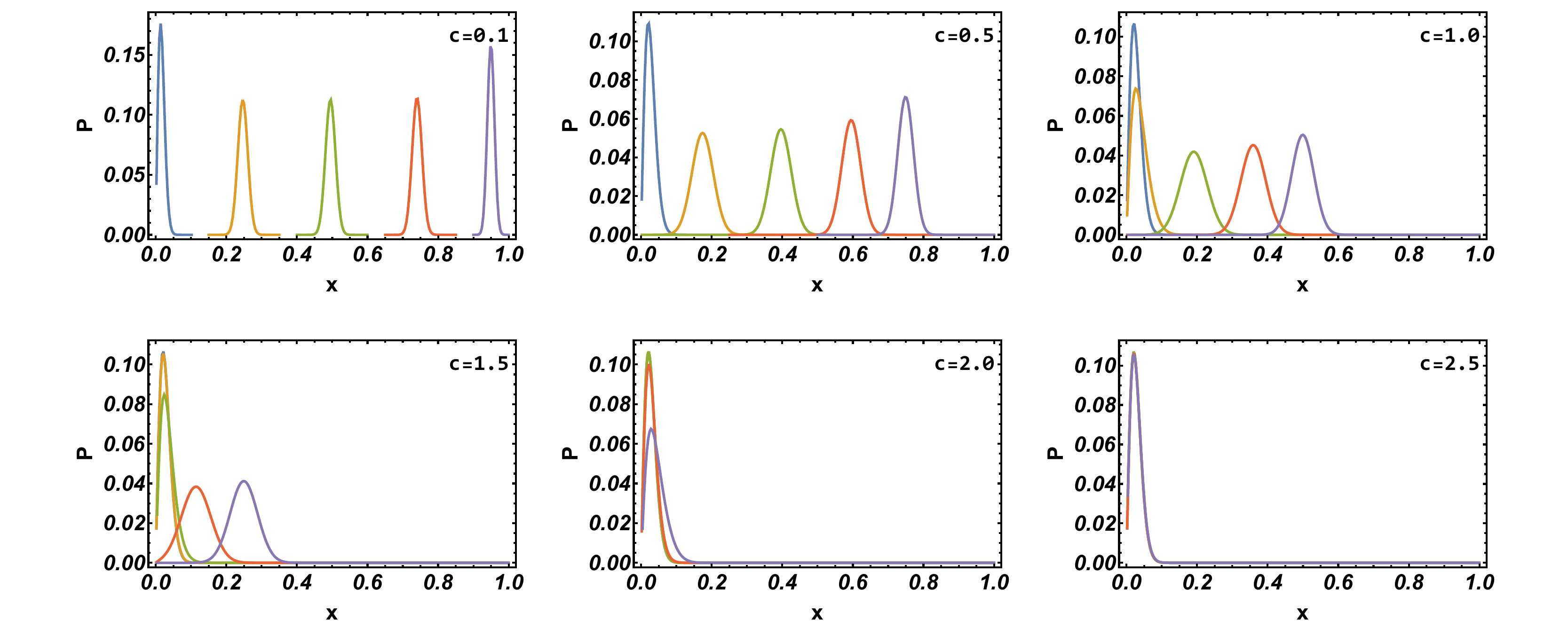}
\caption{Probability density of trajectories of length $t=cK$ on the Krylov chain with hopping amplitudes $b_n=((K-n)/K)^{1/2}$ for different initial conditions. The coloring matches the one in \fig{fig:001}.}
\label{fig:006}
\end{figure}

For $c<c_{\text{cr}}$, the asymptotic Krylov complexity depends on the initial point on the Krylov chain; see Fig.~\ref{fig:001}. In addition, the distribution of fluctuations of the ends is Gaussian, contrary to the KPZ $K^{1/3}$ scaling for $c>c_{\text{cr}}$ case, see \fig{fig:006}. Since effective velocity at late time is  $v\propto c^{-1}$, one could say a bit loosely that the velocity of the fermion is too high and it can not be captured by the effective potential at the ground state.

\begin{figure}[ht]
\includegraphics[width=0.4\textwidth]{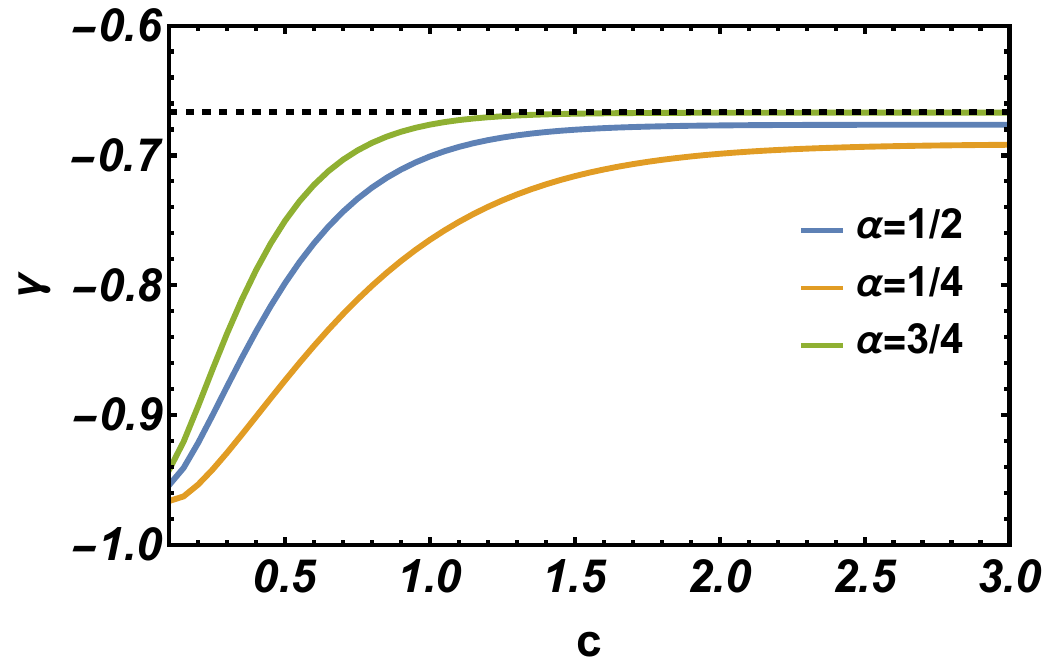}
\caption{Scaling dependence of  the return probability of the random walks of length $t=cK$ on a Krylov chain, where the hopping parameters are the Lanczos coefficients $b_n=(1-n/K)^\alpha$. The black dashed line corresponds to $\gamma=-2/3$.}
\label{fig:002}
\end{figure}

We can consider the autocorrelators $C(1,t)$ that is return probabilities in the Krylov space. The corresponding plots for different $n$-dependencies of the Lanczos coefficients are presented in \fig{fig:002} for the $t=cK$ regime.

\begin{figure}[ht]
\includegraphics[width=0.4\textwidth]{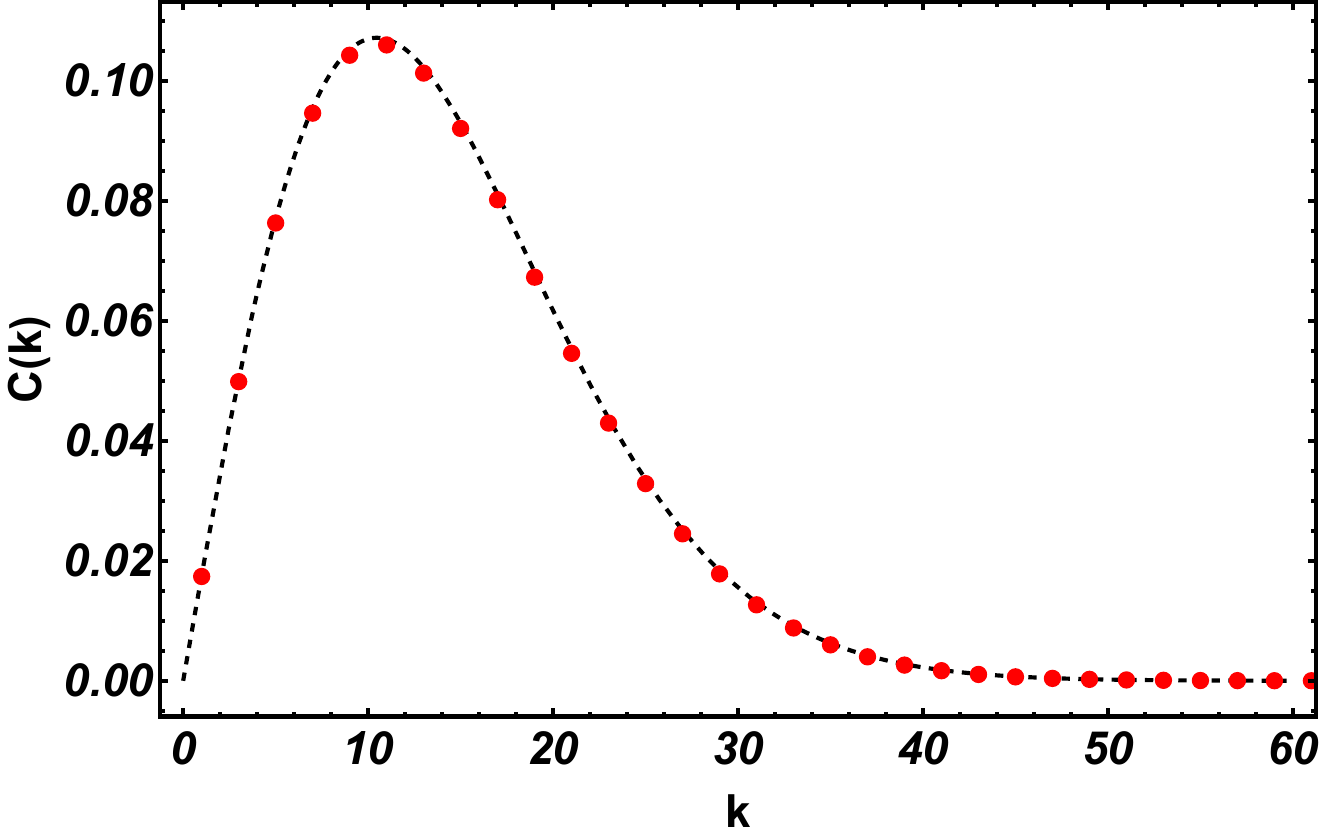}
\caption{Two-point correlator $C(k,t)$ (red dots) and its approximation $2 \alpha^{1/3}K^{-1/3} {\rm Ai}[(2\alpha)^{1/3} K^{-1/3} k+ a_1]$ (black dashed line) for a Krylov chain, where the hopping parameters $b_n=(1-n/K)^{1/2}$.}
\label{fig:007}
\end{figure}

Consider now the two-point correlator and derive the scaling function $f_{kpz}(K^{-2/3}x)$ in terms of the Krylov space. That is, take a look at the two-point correlator $C(r,t)=\la O(0)O_r(t)\ra$ and assume that that site $r$ is not far from the origin and it does not scale with $K$. This is the generalization of the autocorrelator when we consider the time correlation of the seed operator with its descent. We are interested in behavior $C(r,t)$ as a function of both variables. Since the double scaling is assumed, we expect that $C(r,t)= t^{-2/3}f(r,t)$ with some function $f$. The numerical simulation in \fig{fig:007} demonstrates that the Airy function perfectly approximates the two-point correlation function.

\subsubsection{Random Lanczos coefficients}
 
\begin{figure}[ht]
\includegraphics[width=1\textwidth]{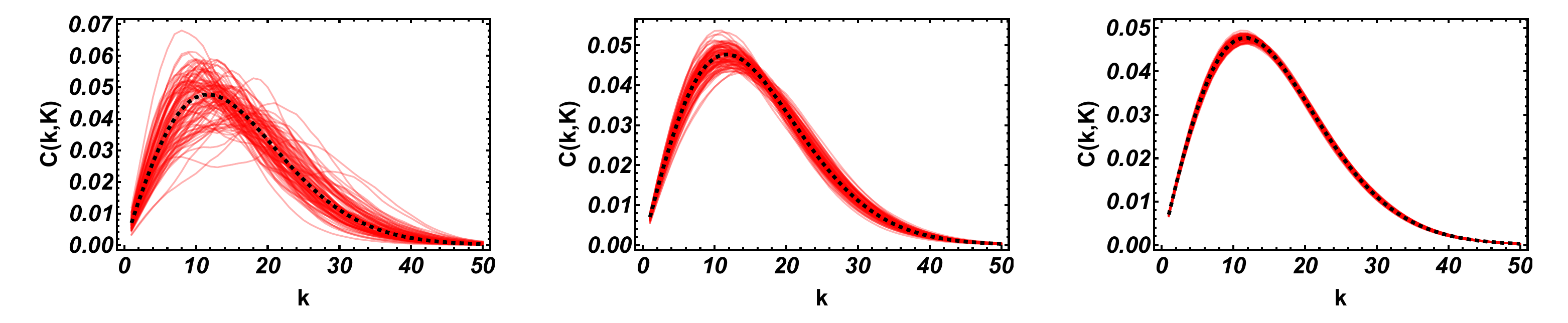}
\caption{Two-point correlators $C(k, K)$ for different values of $\beta$: left panel - $\beta=10$; central - $\beta=100$; right - $\beta=1000$. Black dashed curves correspond to the $\beta=\infty$ case; red curves correspond to an ensemble of $100$ realizations of Krylov chains of length $K=700$.}
\label{Corr_beta.pdf}
\end{figure}

Proceed now with the demonstration that the KPZ-like scaling is well observed at the Heisenberg timescale for random Lanczos coefficients $b_n=(K\beta)^{-1/2}\chi_{(K-n)\beta}$ and $a_n=(K\beta)^{-1/2} \mathcal{N}(0,2)$. First, we examine the behavior of the two-point correlators on Krylov chains with different strengths of disorder represented by $\beta$. We fix the initial point of evolution at the origin and investigate the distribution of the terminal points on the Krylov chain at the Heisenberg time $t\sim K$. In \fig{Corr_beta.pdf}, the probability distribution of endpoints for the ensemble is presented. It turns out that for $t=K$, the fluctuations demonstrate the $\sim K^{1/3}$ scaling.

\begin{figure}[ht]
\includegraphics[width=1.0\textwidth]{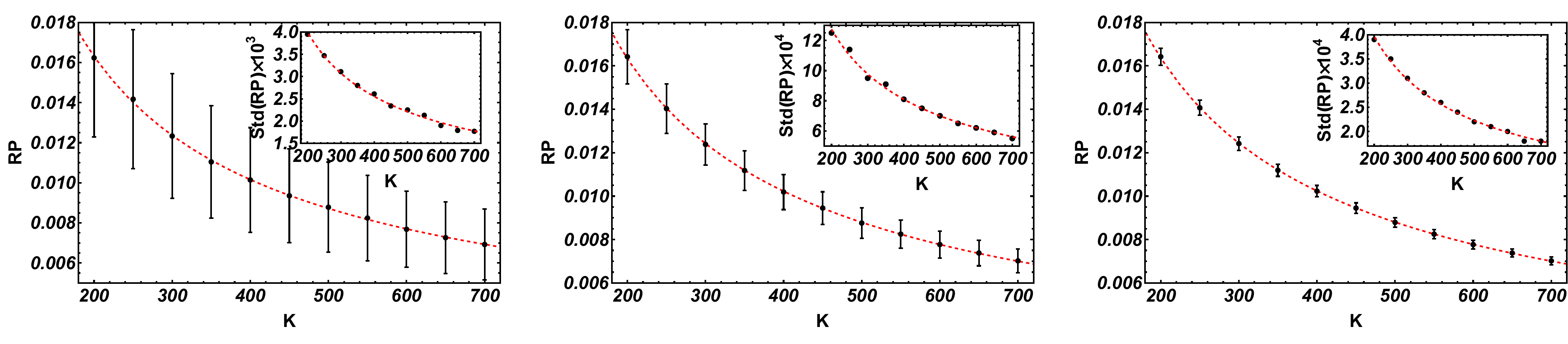}
\caption{Scaling dependence of the return probability $RP=C(1,K)$ averaged over $1000$ realizations of Krylov chains of length $K=700$  for different $\beta$'s: left panel - $\beta=10$; central - $\beta=100$; right - $\beta=1000$. Black dots and error bars represent the mean and standard deviations of the return probabilities; insets show the scaling dependence of the standard deviations. All red dashed curves  are proportional to $K^{-2/3}$.}
\label{RP_beta.pdf}
\end{figure}

Consider now the scaling of return probabilities $RP=C(1, K)$ on the Krylov chain at the Heisenberg time $N=K$ in the plateau regime. We have observed in \fig{RP_beta.pdf} the clear-cut KPZ-like scaling for various disorder strengths represented by $\beta$. The return probability corresponds to the autocorrelator of the seed operator; hence, we predict that the change from Gaussian to KPZ regimes happens quite generically for systems with random Lanczos coefficients. 

Summarizing the results of numerical simulations above, we claim that Gauss-KPZ crossover for descending Lanczos's is clearly seen for one and two-point correlators  at the Heisenberg timescale. The approximate value of the crossover time, $t_{\text{cr}}=c_{cr}K$, has been identified numerically. The Gauss-KPZ crossover happens both for random and deterministic Krylov chains with nontrivial dependence on the noise parameter $\beta$. The crossover time $c_{cr}(\alpha)$ depends on $\alpha$ non-trivially.

\subsection{Growing $b_n$}

\begin{figure}[ht]
\includegraphics[width=0.4\textwidth]{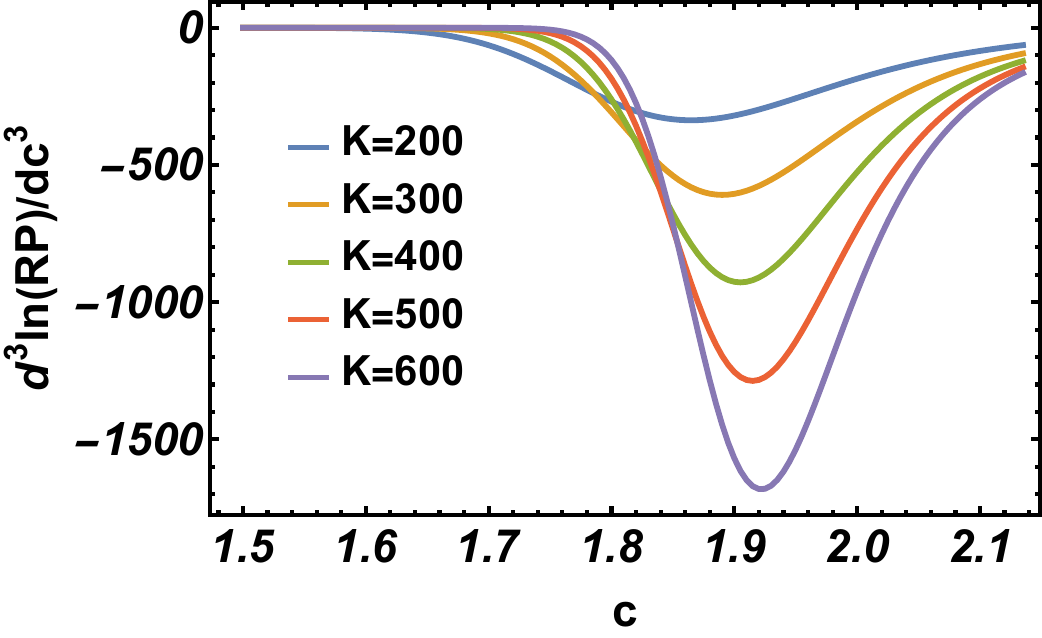}
\caption{The third derivative of the logarithm of the return probability,$\frac{d^3 \ln C(1,cK)}{d c^3}$, on the Krylov chain of size $K$ with growing Lanczos coefficients $b(n)=\sqrt{n}$.}
\label{fig:23}
\end{figure}

\begin{figure}[ht]
\includegraphics[width=0.4\textwidth]{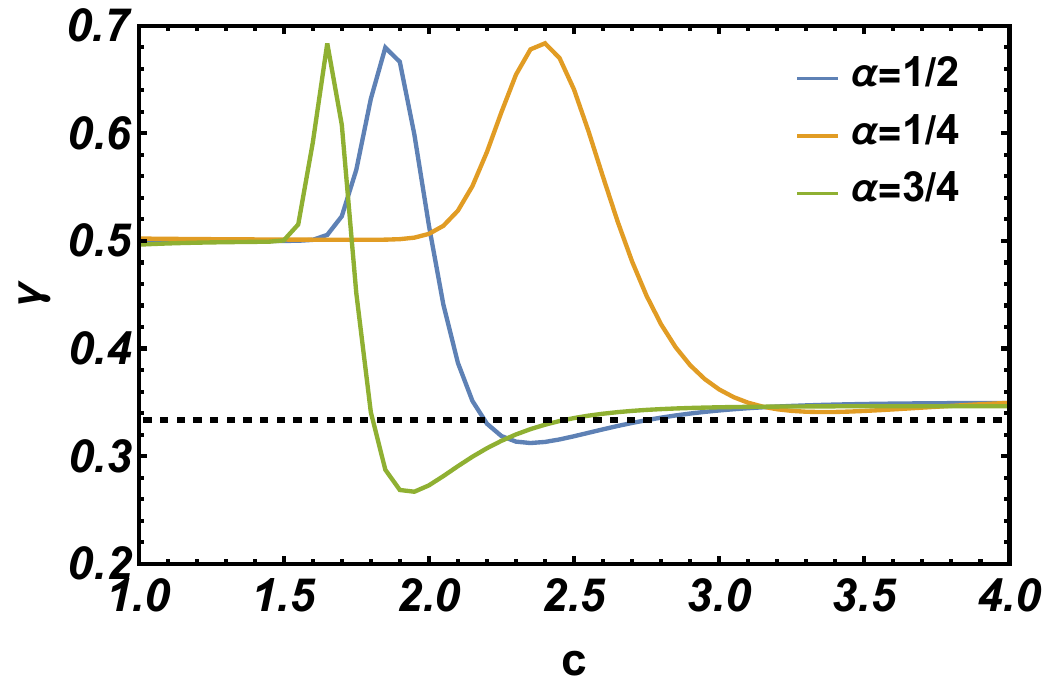}
\caption{Scaling of fluctuations of the middle point of the Brownian bridge of length $2t=2cK$ on a Krylov chain of size $K$, where the hopping parameters are the Lanczos coefficients $b(n)=(n/K)^\alpha$. The black dashed line corresponds to $\gamma=1/3$.}
\label{fig:03}
\end{figure}

Let us now consider a Krylov chain with growing deterministic Lanczos coefficients $b_n=(n/K)^{\alpha}$ and $a_n=0$ and an artificial wall at a distance $K$. Previously \cite{gorsky2018statistical} in related one-body hopping problems on a non-homogeneous chain, it was argued (although not proven rigorously) that we have a 3rd order phase transition at Heisenberg time. To question if we actually have a 3rd order phase transition, let us consider the logarithm of the return probability as the order parameter
\begin{equation}
\log  RP(t)= \log C(1,cK).
\end{equation}
Remarkably, the numerical simulations shown in \fig{fig:23} indeed demonstrate the clear-cut transition in the order parameter at the Heisenberg time. The phase transition at the nearby critical time has been found also for another parameter -- the fluctuations of the middle point of the Brownian bridge of length $2t=2cK$ on a Krylov chain of size $K$ -- see \fig{fig:03}.

\subsection{Full Krylov chain. Summary}

The full Krylov chain for the finite system involves gluing raising and descending regimes
for the Lanczos coefficients. We have analyzed numerically different gluing regimes at different
Euclidean timescales. The summary of our  numerical simulations is as follows:
\begin{itemize}
\item There is no Gauss-KPZ transition at the raising part of $b_n$ dependence;
\item There is the Gauss-KPZ crossover for the descending part of the $b_n$ dependence;
\item The Gauss-KPZ crossover for the full chain exists both for the deterministic and random Lanczos coefficients;
\item The Gauss-KPZ crossover exists for wide interval of powers $\alpha$ in the $b_n$ dependence hence it looks that it holds both for the integrable and chaotic systems.
\end{itemize}

\section{Double scaling in 2D gravity and KPZ scaling via the Krylov space}

\subsection{Matrix models of 2D gravity and ensemble averaging} 

In the previous Sections we have discussed the tridiagonalization of the random matrices
which was considered as the transition to the Krylov basis in some many-body system.
We can question if our findings provide any additional insight on the dual gravity picture 
assuming a kind of holographic picture. Here we comment on the simplest case of the
Gaussian model which corresponds to (2,1) minimal string 
or topological gravity, see \cite{dijkgraaf2018developments} for the review, and focus at the double scaling limit.

The old double scaling limit developed in \cite{ douglas1990strings,brezin1990exactly}, \cite{1990non} served as the tool to zoom the spectral edge region and yields the genus summation in 2D gravity, see \cite{di19952d} for the review.
Consider the Hermitian matrix model with the potential involving the coupling constant $g$
\begin{equation}
{\cal Z}= \int dM e^{V(M)} 
\end{equation}
 say, $V(M)= {\rm Tr}M^2 + g {\rm Tr}M^3$. Pushing the coupling constant to the critical value $g\rightarrow g_{cr}$ where the surface becomes continuous, and imposing the condition $N(g-g_{cr})=\mathrm{const}$, the matrix model variables get defined at the spectral edge where the matrix can be approximated by the differential operator 
\begin{equation}
M \rightarrow -\hbar^2 \frac{d^2}{dx^2} + u(x)
\label{ds}
\end{equation}
and $u(x)$ is the solution to the string equation. In the simplest case of (2,1) gravity one has $u(x)=-x$ and the Airy operator. The Planck constant gets identified with the inverse matrix size $\hbar=\frac{1}{N}$

The natural observables are the averaged over ensemble partition function, 
\be
Z(\rho)=\la {\rm Tr}\, e^{-\rho M} \ra,
\ee
and the spectral formfactor 
\be
Z(\rho\, T)=\la {\rm Tr}\, e^{(-\rho+iT)M} {\rm Tr}\, e^{(-\rho-iT) M} \ra .
\ee
The variable $\rho$ corresponds to the boundary contour length. The multiple correlators $\la Z(\rho_1)\dots Z(\rho_n)\ra$ can be considered as well and they correspond to the surface with multiple boundaries. The behavior of the spectral formfactor at different time scales was used in the investigation of chaotic aspects of 2D gravity. 

Let us push again the system towards the spectral edge, however in a slightly different double scaling limit. As we discussed at length in previous sections, our key object is a sort of an evolution operator $e^{-tM}$ (without the trace) at the late Euclidean time $t$. In this case, the double scaling limit implies $\frac{t}{N}={\rm const}$. In the 2D gravity  this means that we consider the large scale limit $l\sim t$. Once again, we approximate the matrix at the spectral edge by the operator, however we keep the stochastic nature of the matrix which in the simplest Gaussian case yields the stochastic Airy operator at arbitrary value of $\beta$
\begin{equation}
M \to - \frac{d^2}{dx^2} + x + \frac{1}{\sqrt{\beta}} B'(x) = SAO_{\beta}
\label{sao}
\end{equation}
where $B$ is the standard Brownian motion, $B'=\frac{dB}{dx}$ 

Let us make a link with the operator (\ref{ds}) in the double scaling limit of (2,1) minimal string. To this aim let us rescale the $x$-variable and the white noise $B'(x)$:
\be
x = \hbar^{-\mu} y; \quad B'(x=\hbar^{-\mu}y) = \hbar^{\mu/2} B'(y)  
\label{sao2}
\ee
Substituting \eq{sao2} into \eq{sao}, we get:
\be
M\to  -\hbar^{2\mu} \frac{d^2}{dy^2} + \hbar^{-\mu}y + \frac{\hbar^{\mu/2}}{\sqrt{\beta}} B'(y)
\label{sao3}
\ee
To convert \eq{sao3} into the Schrodinger equation, we extract $\hbar^2$ in front of the first term and select $\mu=2/3$:
\be
M\to \hbar^{-2+2\mu}\left(-\hbar^2\frac{d^2}{dy^2} + \hbar^{-\mu+2-2\mu}y + \frac{\hbar^{\mu/2+2-2\mu}}{\sqrt{\beta}} B'(y) \right) = \hbar^{-2/3}\left(-\hbar^2\frac{d^2}{dy^2} + y + \frac{\hbar}{\sqrt{\beta}} B'(y) \right)
\label{sao4}
\ee
Upon general rescaling by $\hbar^{2/3}$ we get the Schrodinger operator which looks as the operator in the 2D gravity with the additional random correction. The random term is suppressed at large $N$ however it can be made $O(1)$ with the proper scaling of $\beta$. 

To recognize the Tracy-Widom distribution consider the eigenvalue equation
\begin{equation}
SAO_{\beta} f= \lambda f
\end{equation}
and take the logarithmic derivative $W=d \log f$. We obtain for SAO the Langevin equation
\begin{equation}
dW= \tfrac{2}{\sqrt{\beta}}db + (t-\lambda -W^2)dt
\end{equation}
and the corresponding Fokker-Planck (FP) equation reads
\begin{equation}
\partial_t F(t,x) + \tfrac{2}{\beta}\partial_{xx} F(t,x) + (t-x^2)\partial_x F_{\beta}(t,x)=0.
\end{equation}
The following limit of the solution to the FP equation 
\begin{equation}
{\cal F}(t)= \lim_{x\rightarrow \infty}F_{\beta}(x,t)
\end{equation}
yields the Tracy-Widom distribution $TW_{\beta}$. It generalizes the Airy distribution when the stochastic term is suppressed at $\beta=\infty$.

Let us comment  on the holographic representation of the 2D gravity via matrix models. This approach worked well in the old matrix models \cite{di19952d} and recently the matrix model were found for the JT gravity \cite{saad2019jt} where it was identified as the minimal string -- the Liouville theory coupled to $(2,p)$
minimal model \cite{maldacena2004exact} at $p\rightarrow \infty$. 
The holographic matrix model realization of 2D gravity raises the question regarding its representation as ensemble average \cite{saad2019jt} and the role of non-perturbative effects in factorization problem \cite{saad2021wormholes}. The single matrix in the matrix model is considered as the random Hamiltonian in the boundary theory. On the other hand in the standard double scaling limit
we get 2d gravity without randomness. The additional random term in the double scaling limit provides the 
hint for the  partial resolution of the tension concerning the averaging issue.

The Tracy-Widom distribution has been discussed in the context of holographic matrix model representation of the JP gravity \cite{johnson2022microstate}. It was argued there that TW distribution can be responsible for the account of the non-perturbative effects due to the discrete nature of the spectrum and for the interpretation of the discreteness of the spectrum in the Lorentzian and Euclidean viewpoints. Non-perturbative wormhole configurations have to be taken into account at the ramp-plateau transition. Our study provides even more natural pattern for the emergence of the TW distribution: we claim that presumably it can be derived upon  a more careful representation of the random matrix at the spectral edge.

Note that very recently \cite{altland2024quantum} it was suggested that the behavior of the system at the spectral edge can be considered as a kind of the quantum  phase transition. To this aim the control parameter gets identified with energy, while the order parameter is the spectral density which undergoes the non-analytic transition at the spectral edge: $\rho(\lambda) \sim \sqrt{\lambda}$. It was argued that one can distinguish two different phases: sparse and dense, which have the same non-analyticity but are described by the different effective theories. The physical origin of the difference between two regimes is the different behaviour of fluctuations, which develop a kind of collective behavior in dense case. 

With these new points we suggest the identification of the KPZ scaling for gravity observables assuming that effective ``Euclidean time variable'' is the boundary length. Hence we can search for the KPZ scaling at the asymptotically large boundary lengths. In the JT gravity context it is the region near the $AdS_2$ boundary, and the boundary length can be written in terms of the boundary value of  dilaton $l=\phi$.

\subsection{The noise strength as the central charge}

We have seen above that zooming at the spectral edge in the double scaling limit in the Gaussian $\beta$-ensemble we get the stochastic Airy operator and $\beta$ fixes the noise strength. Let us point out that similar interpretation can be derived even without passing to the spectral edge. Recall that the probability 
\begin{equation}
P_{\beta}(\lambda_i) = \prod_{i<j} (\lambda_i -\lambda_j)^{\beta}e^{-\beta \lambda_i^2}
\end{equation}
can be considered as the norm of the wave function for the rational Calogero model supplemented by the oscillator potential. There is a duality \cite{nekrasov1997duality} between the rational Calogero model in the harmonic trap and the trigonometric Calogero model on the circle of circumference $R$ with Hamiltonian 
\begin{equation}
H_{tr}= \sum_i p_i^2 + \beta(\beta-1)\sum_{i< j} \frac{1}{\sin^2((x_i -x_j)/R)^2}
\end{equation}
The radius of the circle in this case is $R=\frac{1}{\beta}$ and the very duality has the meaning of transition to the ``radial coordinates". The Hamiltonian in one system gets mapped into the momentum in the dual.

Now turn to the relation between the trigonometric Calogero-Moser system and multiple radial SLE \cite{cardy2003stochastic}. One considers the set of points at the boundary of the disc representing the interacting Calogero particles. These paints are the ends of  growing ``strings" reaching the center of the disc where the operator $V(0)$ is inserted. The growing process occurs in the noisy background and is described by the
radial SLE process. To formulate the evolution let us first consider the infinitesimal conformal transform $z\rightarrow z+\alpha(z)$ where $\alpha(z)= \sum_{j=1}^{N} b_j \alpha_j$
\begin{equation}
\alpha_j(z) = -z \frac{z+ e^{i\theta_j}}{z- e^{i\theta_j}}
\end{equation}
$b_j$ are infinitesimal parameters and $\theta_(t)$ are time dependent variables living at the disc boundary.
Assume that they evolve according to the Brownian motion
\begin{equation}
d\theta_j = \sum_{k< j}\cot\frac{(\theta_j-\theta_k)}{2} + \frac{1}{\sqrt{\beta}}dB_j(t)
\end{equation}
where $B_{j}(t)$ are independent Brownian motions. This evolution describes the multiple radial SLE growing process when the points are at the boundary at $t=0$ and reach the center of the disc at $t=\infty$. It is interesting to consider the asymptotic equilibrium solution to the corresponding Fokker-Planck equation for the joint probability distribution $P([\theta_i],t)$. It coincides up to gauge conjugation with the eigenvalues of the trigonometric Calogero-Moser system \cite{cardy2003stochastic}.

\begin{figure}[ht]
\centering
\includegraphics[width=0.3\linewidth]{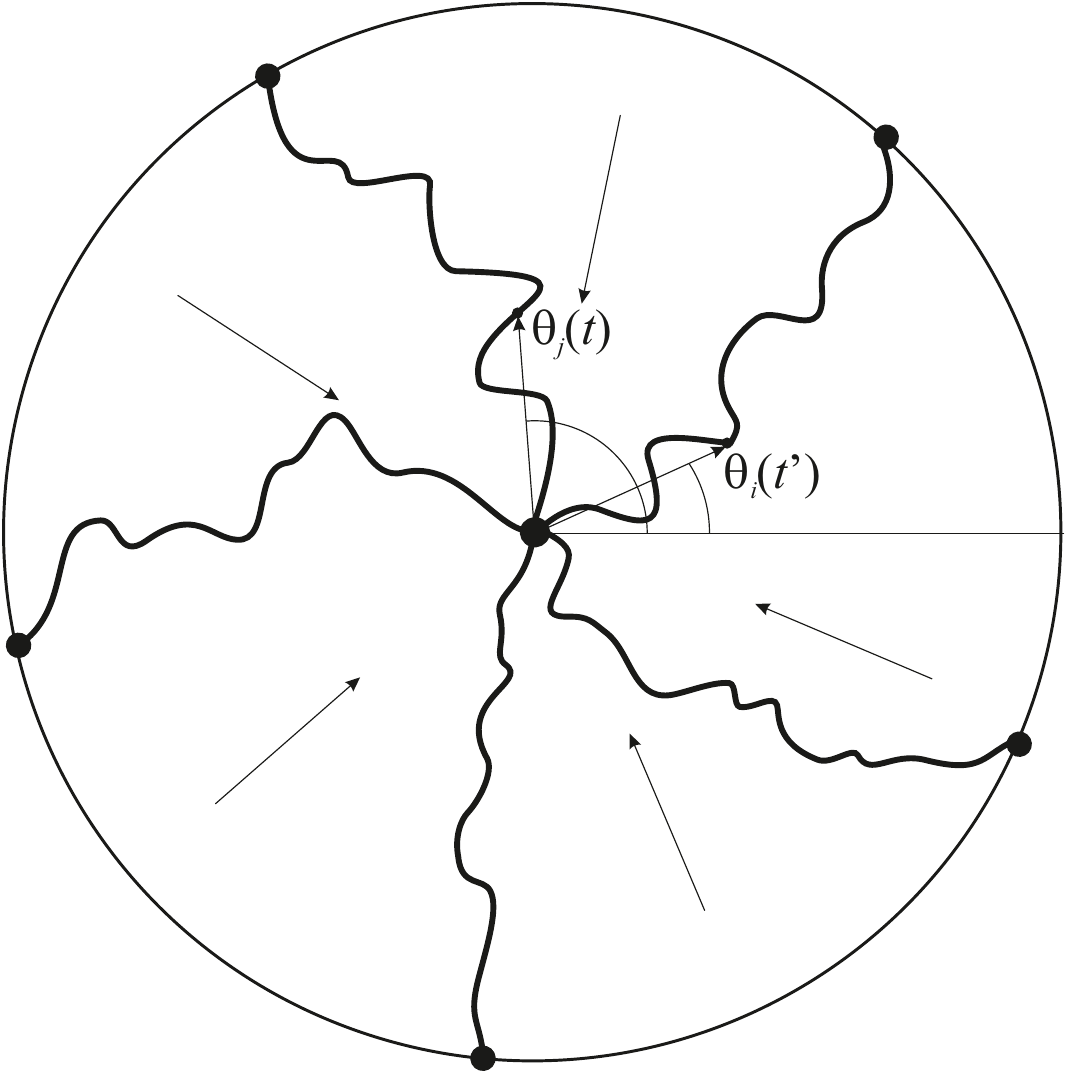}
\caption{Multiple radial SLE process. Calogero degrees of freedom live at the disc boundary.}
\label{fig:krylov-sle2}
\label{krylov-sle2}
\end{figure}

As in the stochastic Airy operator at the spectral edge discussed in the previous Sections here the 
parameter $\beta$ fixes the strength of the noise. Moreover it also fixes the central charge of the auxiliary conformal theory on the unit disc with conformal boundary conditions
\begin{equation}
c= 1- 6\left(\sqrt{\beta} - \frac{1}{\sqrt{\beta}}\right)^2
\end{equation}
The Calogero-Moser wave function can be represented as the correlation function
\begin{equation}
\Psi(z_1, \dots z_N)\sim  \la \phi_{2,1}(\theta_1)\dots \phi_{2,1}(\theta_N)V(0)\ra
\end{equation}
where the primary field $V$ is inserted at the origin where the ``strings'' in the radial SLE end up, see
\fig{fig:krylov-sle2}.
The highest weight of $V$ is related to the eigenvalue value of the Calogero Hamiltonian.

\section{Discussion}
\subsection{KPZ scaling in spin chains}

We have already pointed out that the results obtained in \cite{ljubotina2019kardar, scheie2021detection, bulchandani2021superdiffusion, ilievski2021superuniversality, de2020superdiffusion, PhysRevE.100.042116, krajnik2020kardar} motivated our study. The superdiffusive anomalous transport has been found numerically in the late-time two-point correlators for the integrable Heisenberg spin chains at half filling
\begin{equation}
\la S^z(x,t)S^z(0,0) \ra \propto (\Gamma t)^{-\frac{2}{3}} f_{kpz}\left((\Gamma t)^{-\frac{2}{3}}x\right)
\end{equation}
where $\Gamma$ is parameter of the model and $f_{kpz}$ is tabulated function. The same scaling exponent has been found for autocorrelators of the  operators
\begin{equation}
\la S_i^z(t)S_i^z(0) \ra \propto (\Gamma t)^{-\frac{2}{3}} 
\end{equation}

It was observed later that the KPZ scaling survives upon adding the integrability breaking terms that is the integrability is not the origin of the phenomenon. Besides, it was argued that the presence of the global symmetry is important. Presumably in the integrable situation the origin of the KPZ scaling exponent in the integrable case is the formation of the long Bethe strings and the decay of the short strings nearby the long ones. More recently the presence of universal superdiffusion in correlators has been found in quantum gases \cite{wei2022quantum}, superconducting qubits \cite{rosenberg2023dynamics} and neutron systems \cite{scheie2021detection}.

Saying differently, it has been suggested that the giant soliton is formed and the short strings provide its 
renormalization. The soliton was attributed either to the ``soft gauge node'' in the effective description \cite{bulchandani2020kardar} or some effect of interactions in the generalized hydrodynamics. In both interpretations the observed anomalous spin diffusion in the Heisenberg chain follows from interacting long-wavelength spin fluctuations. The exact identification of the giant soliton in the generalized hydrodynamics with the classical soliton of the Landau-Lifshitz equation has been found in \cite{de2020superdiffusion}.

The KPZ equation can be rewritten as the stochastic Burgers equation which provides quite qualitative picture behind the  $z = 3/2$, $t=x^z$ scaling exponent for fluctuations which can be derived as follows. At equilibrium, the typical value of Burgers field $u(x,t)$, coarse-grained over a region of size $x$, scales as $x^{-1/2}$. This is also the characteristic velocity at which fluctuations traverse this region. Therefore, the time taken for a fluctuation to traverse a region of size $x$ is $t(x) \propto \frac{x}{x^{-1/2}}\propto x^{3/2}$. However these arguments seem too naive and more rigorous consideration based on the underlying universal algebraic structure and the Yangian symmetry has been recently formulated in \cite{ilievski2021superuniversality}. 

How the Gauss-KPZ crossover we have found for the finite systems  could help in explanation of transition to KPZ regime for the spin chains? The most evident outcome follows for the autocorrelator of the spin operators at the same site. We have found in quite general setting for Lanczos coefficients that the autocorrelator at the Heisenberg timescale for the finite system enjoys the KPZ-like scaling. Selecting the operator $S_i^z$ as the seed operator, we get the  finite Krylov chain. 
We can expect the following behavior of the two-point correlator of the local seed operator and the non-local operator at $k$-th level of the finite Krylov chain of length $K$ at the Heisenberg time $t\sim K$:
\be
\la S_i^z(0) O_k(t) \ra \sim K^{-1/3} {\rm Ai}(K^{-1/3}k)
\ee
  It was observed recently numerically that for non-integrable case the inverse superdiffusion-diffusion crossover occurs even at larger time \cite{mccarthy2024slow}. This transition has been attributed to the decay of the giant soliton. We do not observe such inverse transition in our models of Lanczos coefficients but can not exclude that the time in numerical simulations is not large enough.
 
The most interesting question concerns the evaluation of the space-time correlators of local operators $\la S_i^z(0) S_j^z(t) \ra$ however it is beyond the scope of our study and we postpone this issue for the separate work restricting ourselves by  few general remarks. Clearly the first step should be the incorporation
the space coordinate $x$ in the Krylov space approach which naively does not know about it. For any seed operator we can get $O(x)= e^{iPx}O(0)e^{-Px}$ where $P$ is momentum operator in the particular model. In the translationally invariant model we can develop the Lanczos algorithm for the operator $P$ similar to the Liouvillian which would yield the second Krylov chain with the hopping Lanczos coefficients $\tilde{b}_n, \tilde{a}_n$. The Lanczos coefficients for Hamiltonian and momentum Krylov chains are different. One can also consider the instant correlations $C(x)=\la O|e^{iPx}|O\ra$ similarly to the autocorrelators. 

In the translationally invariant models Hamiltonian and momentum commute hence we can form a kind of  Krylov lattice from two Krylov chains. The generic words formed from the $H,P$ operators  correspond to the different paths on the Krylov lattice. However in the interacting models with disorder the translation invariance can be  broken and the derivation of the Krylov lattice from the pair of the Krylov chain  seems to depend on the details of the interaction.

\subsection{Is there the DQPT for the dynamics on the Krylov chain at critical time?}

We have observed  the clear-cut fingerprints of Gauss-KPZ 3-rd order transition which happens at the Heisenberg Euclidean timescale for the growing Krylov chain with cut-off.  One could question if it can be interpreted  as the DQPT which takes place in some out-of-equilibrium systems, see \cite{heyl2018dynamical, marino2022dynamical,zvyagin2016dynamical} for the reviews. The DQPT occurs at a peculiar Minkowski time when some observable behaves non-analytically. The standard Landau-Ginzburg-like approach based on the mean-field approach for some effective field whose vev serves as the order parameter is not working. Hence the identification of the order parameter for DQPT is not a simple issue and a Loschmidt echo is sometimes used for this aim. In our case the we work with  a Euclidean time hence it is not strictly DQPT but could bear some of its features.

The interesting example of the 3rd order DQPT has been found for the late-time dynamics in the XY spin model with some boundary conditions \cite{perez2024dynamical}. The authors have used the known relation of the transition amplitude in the XY model and the partition function of the unitary matrix model. It is well known that there is the Gross-Witten-Wadia (GWW) 3rd order phase transition in this model \cite{gross1980possible,wadia1980n} when the spectral density gets reorganized. The transparent  viewpoint at GWW transition due to the complex saddles has been developed in \cite{buividovich2016complex}. The logarithm of the return probability has been used in \cite{perez2024dynamical} as the order parameter and it was demonstrated both analytically and numerically  that the model undergoes the 3rd order phase transition at some critical time $T=\tau_{cr}N$ where $N$ is the system size. From the GWW model viewpoint the critical point corresponds to the fixed value of t'Hooft coupling. The $\tau_{cr}$ has been also linked in \cite{perez2024dynamical} to the quantum speed limit.

The microscopic generic mechanism behind the DQPT is not completely clear. It is believed now (and was checked in several examples) that the production and condensation of the topological defects with a non-vanishing density matters \cite{flaschner2018observation}. The discussion of the 3rd order DQPT transition in \cite{perez2024dynamical} provides new support for this interpretation. Indeed the 3rd order Douglas-Kazakov phase transitions \cite{douglas1993large} for 2D YM or 2D qYM were clearly identified with the condensation of instantons \cite{gross1995some, PhysRevD.73.026005}. Similarly, the 3rd order transition can be mapped on the condensation of windings for vicious walkers \cite{gorsky2020two,forrester2011non}. There is some similarity of this phenomena with the formation of superfluid component \cite{gorsky2020two}. The review of 3rd order phase transitions can be found in \cite{majumdar2014top}. 
Our findings bear some similarity with  interpretation of the microscopic physics at the Heisenberg time as the condensation phenomenon. Indeed, the transition occurs when there is enough time for the probe particle to explore the whole finite-dimensional Krylov chain corresponding to some seed operator. That is we have a possible ``winding'' around the Krylov chain. However  the additional study is required to justify this interpretation.

\section{Conclusion}

In this study we considered the universal behavior of the late-time operator dynamics in systems with finite-dimensional Hilbert space in the peculiar double scaling limit. It corresponds to the Heisenberg timescale when the evolution time is of order of the Krylov chain length for the selected seed operator. The Krylov complexity starts to saturate around the Heisenberg timescale while the Lanczos coefficients
change the growing regime to descending regime.

We have found in quite general setting that the diffusion-superdiffusion or Gauss-KPZ transition is the crossover which occurs at the Heisenberg timescale at the descending part of $b(n)$. We identify numerically that above the critical time the late time correlators and autocorrelators  described as 
transition amplitudes along the Krylov chain manifest the KPZ-like scaling exponents: the $K^{1/3}$ fluctuations law for the one-point functions and the $K^{-2/3}$ scaling for autocorrelators. The transition to superdiffusion has been found both for the deterministic and random Lanczos coefficients. 
Our finding provides the possible explanation of late time KPZ scaling in the spin chain correlators and suggests that the similar behaviour can be expected for  other systems with finite Hilbert space
at the Heisenberg timescale. Let us emphasize that we focused at the Heisenberg timescale $t\sim K$ and the investigation of the regimes $t\sim K^m, m>1$ deserves the additional study.

We have also considered the case of the purely growing Lanczos coefficients with additional cut-off which makes the Krylov chain finite. In this case the Gauss-KPZ transition turns out to be 3rd order phase transition. For some specific value of disorder the analytic results concerning the stochastic Airy operator from the probability theory provide  rigorous ground for our numerics.
We have  discussed numerically the case of the stick seed operator which corresponds to the non-vanishing $a_0$ Lanczos coefficient. We have found that there is a kind of criticality at some values of $a_0$ similar to discussion in \cite{gorsky2023unconventional}. It would be interesting to investigate the effect of more general non-vanishing coefficients $a_k$.

Our study provides some additional tools for the analysis of non-perturbative effects in 2D gravity which are important in the double scaling when in the dual matrix model near the spectral edge the matrices are approximated by the differential operators. It was argued in \cite{johnson2022microstate} that
the discreteness of the spectrum matters and could lead to the Tracy-Widom distribution. At the gravity side it is assumed that wormholes contribute the non-perturbative effects. We have argued that Tracy-Widom distribution also emerges in another way taking into account that the differential operator has the random term which yields the KPZ scaling at small effective temperatures corresponding to large boundary lengths in the matrix model macroscopic loop. For instance in Krylov basis the individual matrix from the Gaussian ensemble near the spectral edge is approximated not by the Airy differential operator but SAO which yields the clear-cut link to the KPZ scaling at the late-time evolution. 

Since the noise term in the operator indicates the presence of the auxiliary CFT whose central charge governs the noise strength it is desirable to identify properly the gravity meaning of this auxiliary CFT and corresponding semiclassical limit at $\beta=\infty$. The related question concerns the modification of the KdV hierarchy familiar in the double scaling limit of the matrix model due to a noisy terms in the Lax operator at finite $\beta$. Another important question is to identify the microscopic
configurations, presumably wormholes, responsible for the noise term.

The noisy term in the Lax differential operator in the double scaled limit raises the natural question concerning the localization properties of its eigenfunctions. In the Krylov basis several regimes including a bit exotic non-ergodic extended  phase have been identified in \cite{das2024robust}
at particular scaling of $\beta$ and it would be important to identify properly the gravitational aspects of the Anderson localization with diagonal disorder in terms of  the $SAO_{\beta}$. We shall discuss these issues elsewhere.

There are many other open questions related to our study and one of the most pressing one concerns the evaluation of the space-time correlators in the Krylov basis framework. The derivation of the full KPZ function $f_{KPZ}(t^{-2/3}x)$ in the Krylov space approach is much more subtle since in this case  it is not enough to consider the single Krylov chain. We have conjectured  that the second Krylov chain gets formed for the momentum operator similar to the Krylov chain for Liouvillian. Together the Hamiltonian and momentum chains form a kind of the Krylov lattice and the full correlator can be describe via the transition amplitude of the probe particle on this lattice with the directed links. We have no general solution in this case yet and hope to discuss this issue elsewhere.

Our parameter $c$ plays the role of the ``inverse velocity'' for the late-time motion of the probe particle on the Krylov chain. The criticality in $c$ we have found suggests that it could be related with the quantum speed limit that generalizes the energy-time Heisenberg and Robinson uncertainty relations. It quantifies the shortest time to get the final state from the initial one.  The adaptation of the quantum speed limit for the Krylov chain has been done in \cite{hornedal2022ultimate, carabba2022quantum}. In this case once again we consider the orthogonalization at each step but instead of the energy level the proper variable is the Krylov complexity. It would be interesting to discuss the possible link of our findings with this issue.

Another interesting question concerns the possible relation with the Lifshitz tail in the spectrum of disordered systems. In some non-rigorous sense it can be considered as Laplace dual of the KPZ scaling \cite{gorsky2021lifshitz}. Holographically the 
Lifshitz tail corresponds to the behavior of the random curves in $AdS_2$ in the UV and it would be interesting to make the link of this picture with our study.

We are grateful to A. Alexandrov, A. Artemev, A. Litvinov and B. Meerson for the useful discussions. A.G. thanks IHES where 
the part of the work has been done for the hospitality and support.

\begin{appendix}

\section{Krylov chain with a sticky seed operator}

Here we shall consider a bit more general situation and assume that the Lanczos coefficients $a_0\neq 0$. In terms of the Krylov basis it is defined as 
\begin{equation}
a_0=\la O_0|{\cal L}|O_0 \ra
\end{equation}
and can be considered as the intrinsic characteristic of the seed operator. Let us emphasize that its value can be different at the different regions in the Hilbert space. Below we are investigating numerically the dependence on $a_0$ both for random and deterministic systems and find for some observable a kind of a critical behavior.

It is instructive to study fluctuations of the midpoint of the Brownian bridge (Bb) on a Krylov chain $k=1,2,..., K$ ($K\to \infty$) with a sticky boundary at $k=1$.  To be precise, consider the following scaling of hopping rates, $b_k$:
\be
b_k= (K-k)^{\alpha} = \begin{cases} 1  & \mbox{uniform amplitudes, $\alpha = 0$} \medskip \\ 
\sqrt{K-k} & \mbox{descending amplitudes, $\alpha=1/2$} \end{cases}
\label{eq:hopping}
\ee
where $K$ is the lattice size. 

We search for the transition point controlled by the depth of the potential well (the ``stickiness''), $a_0$, expecting that the order of the phase transition depends on the scaling exponent $\alpha$ of the hopping amplitude. The statistics of a Bb on a 1D Krylov lattice (in discrete time) is determined by the conditional probability $Q_N(k,n|K)$ to find the $n$'th step of a Bb at the distance $k$ from the boundary:
\be
Q_N(k,n|K)=\frac{P_n(k)P_{N-n}(k)}{\sum\limits_{k=1}^K P_n(k)P_{N-n}(k)}=\frac{1}{\cal N}[{\cal L}^n]_{k,1}[{\cal L}^{N-n}]_{k,1}
\label{eq22}
\ee
where $P_N(k)$ is the probability to find the end of the open path of length $N$ at the distance $k$, $[{\cal L}^N]_{k,1}$ is the element $(k,1)$ of the matrix ${\cal L}^N$, and ${\cal N}$ is the normalization constant for the conditional distribution $Q_N(k,n|K)$. In what follows without the loss of generality and for simplicity we take $n=N/2$, i.e. we consider the distribution of the Brownian bridge midpoint. 

\begin{figure}[ht]
\includegraphics[width=0.45\textwidth]{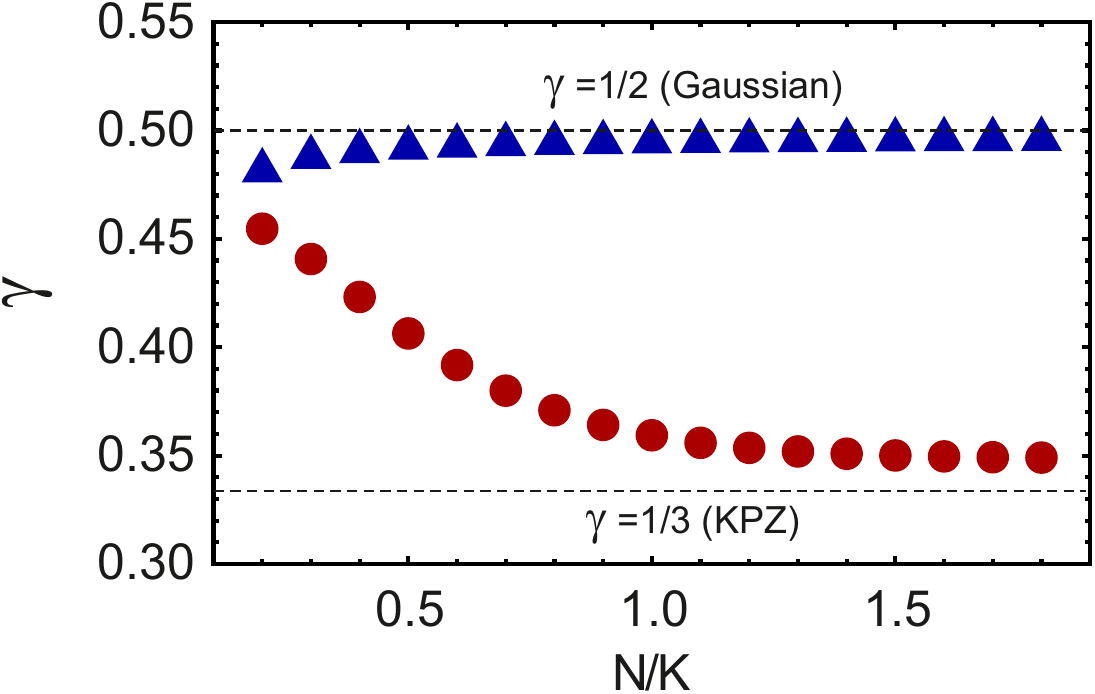}
\caption{Limiting behavior of critical exponent $\gamma$  of Brownian bridge in a double-scaling regime ($N\to\infty$, $K\to\infty$, $N/K={\rm const}$): Blue triangles -- for the uniform hopping rates ($\alpha=0$), red circles -- for the descending hopping rates ($\alpha=1/2$).}
\label{fig:05}
\end{figure}

Equation \eq{eq22} permits us to compute the scaling of a typical span $\Delta(N) \sim N^{\gamma}$ of a Brownian bridge on a 1D lattice with {\it descending} hopping amplitudes in the doubly scaling limit $N\to\infty$, $K\to\infty$, $N/K={\rm const}$ and find the critical exponent $\gamma$, which coincides with the KPZ growth exponent, $\gamma_{KPZ}=1/3$. To the contrary, the critical exponent $\gamma$ for the Bb on a lattice with {\it uniform} transition rates has been repeatedly computed (see, for example, \cite{khokhlov}) where it has been shown that such a Bb has a span $\Delta$ controlled by the exponent $\gamma_{Gauss}=1/2$. The plots demonstrating the saturation of the exponent $\gamma$ for the lattices with uniform and descending amplitudes are shown in \fig{fig:05}. 

\begin{figure}[ht]
\includegraphics[width=0.8\textwidth]{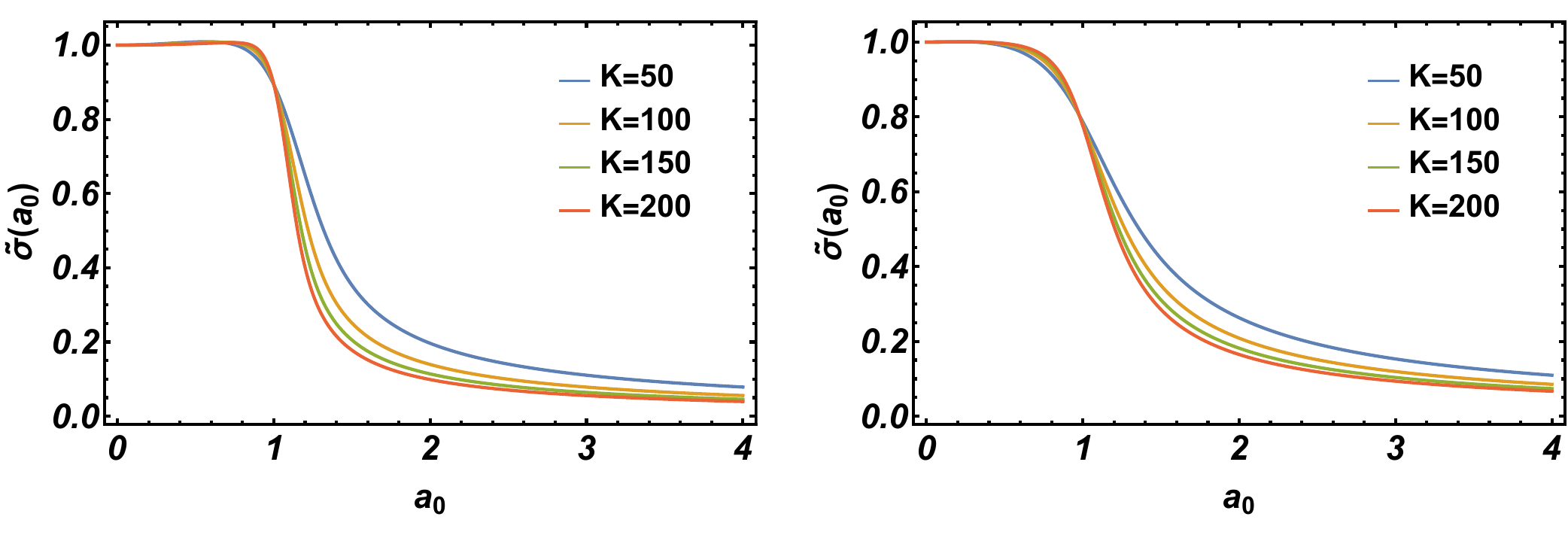}
\caption{Dependence of the rescaled variance, $\tilde{\sigma}$ on $a_0$ for a few values of $K$: (a) Uniform hopping amplitudes ($\alpha=0$); (b) Descending hopping amplitudes ($\alpha=1/2$).}
\label{fig:06}
\end{figure}

To determine numerically the order of the localization transition of a Bb on the 1D lattice with a sticky boundary, we proceed as follows. First we define the dependence of the rescaled variance, $\tilde{\sigma}(a_0)=\sigma(a_0)/\sigma(a_0=0)$, of the Brownian bridge midpoint on the stickiness $a_0$ for two different kinds of hopping amplitudes (uniform and descending) and for various lattice sizes $K$. The corresponding plots are shown in \fig{fig:06}a,b.

\begin{figure}[ht]
\centering
\includegraphics[width=0.8\textwidth]{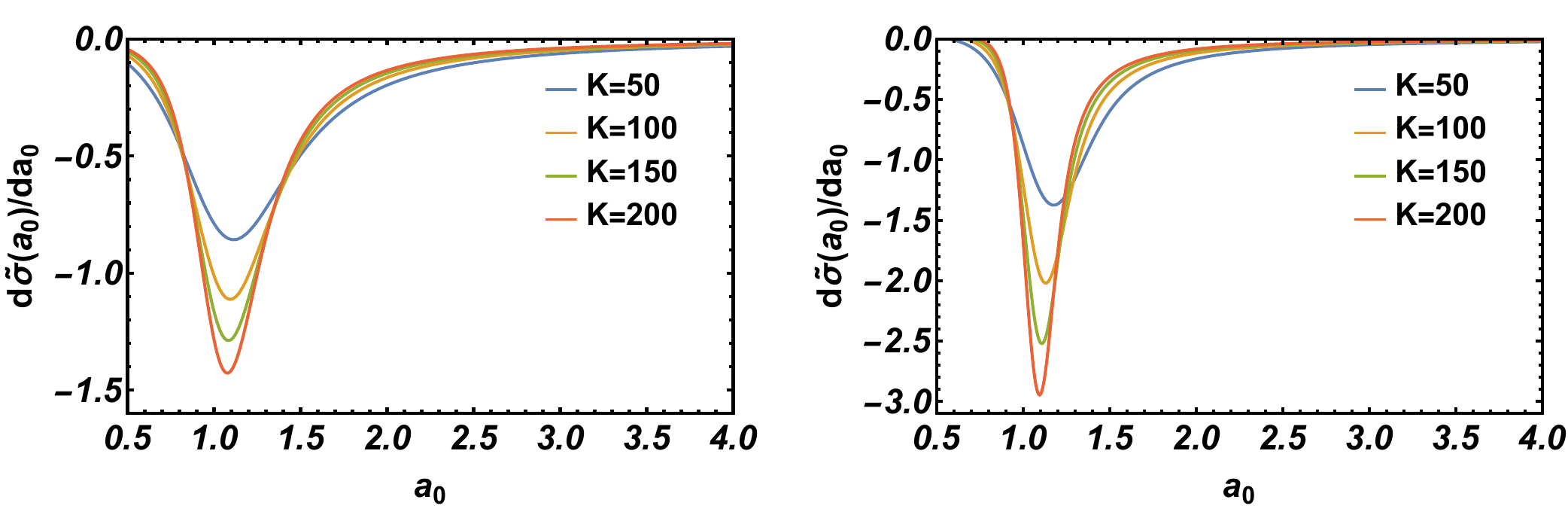}
\caption{Plot of the function $\tilde{\sigma}'(a_0)$: (a) Uniform hopping amplitudes ($\alpha=0$); (b) Descending hopping amplitudes ($\alpha=1/2$).}
\label{fig:07}
\end{figure}

Next, we numerically compute the derivative $\tilde{\sigma}'(a_0)\equiv \frac{d\tilde{\sigma}(a_0)}{d a_0}$ and associate the transition width, $\Delta$, with the width of the function $\tilde{\sigma}'(a_0)$ at the level $\min(\sigma(a_0))/\sqrt{2}$ for every $K$ -- see \fig{fig:07}a,b for regular and descending hopping amplitudes.

\begin{figure}[ht]
\centering
\includegraphics[width=0.45\textwidth]{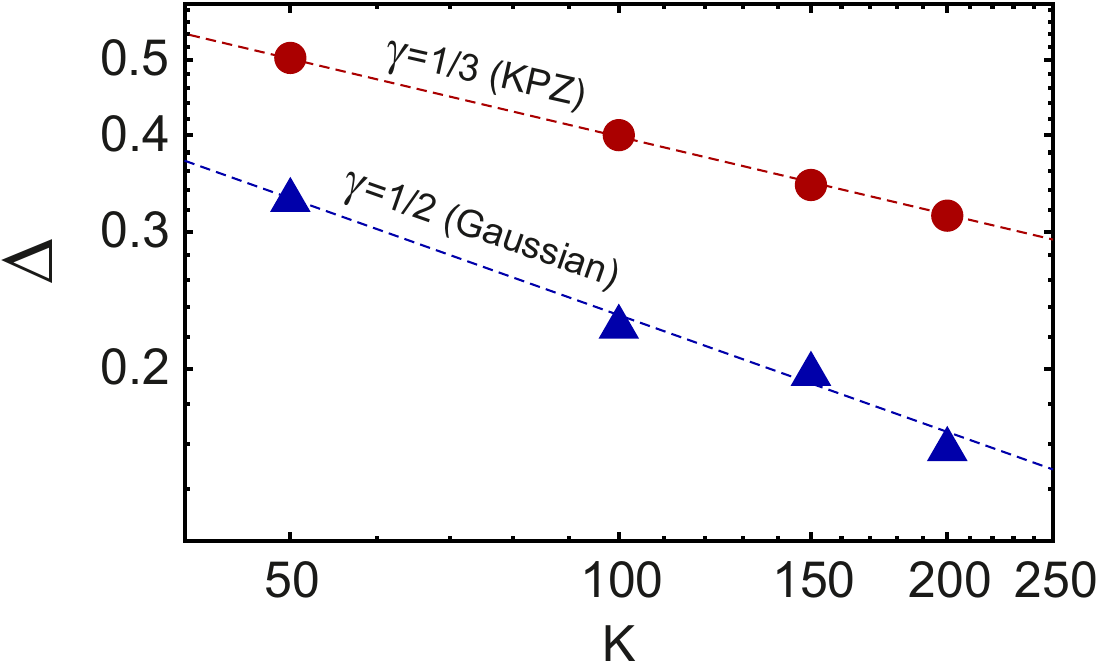}
\caption{The finite-size scaling (in a log-log scale) of the width of adsorption transition as a function of a size of a tree: the Gaussian exponent $\gamma=1/2$ for uniform amplitudes (blue triangles); the KPZ exponent $\gamma=1/3$ for descending amplitudes (red circles).}
\label{fig:08}
\end{figure}

It is well known \cite{binder1987finite} that the order of the phase transition, $\theta$, can be extracted from the finite-size dependence of the transition width, $\Delta$ on $K$. Namely, if $\Delta$ shrinks with $K$ as $\Delta\sim K^{-1/\theta}$, then the transition order at $K\to\infty$ is $\theta$. The dependence of the width $\Delta$ on the lattice size, $K$, together with the power-law approximation $a K^{-1/\theta}$, are plotted in \fig{fig:08} in doubly-logarithmic coordinates.

For both rates $b_k$, uniform ($\alpha=0$) and descending ($\alpha=1/2$), the localization transition width, $\Delta$, behaves as $\Delta \propto K^{-1/\theta}$ where $\theta=D_f$ and $D_f$ is the fractal dimension of Brownian bridge on the corresponding lattice. From plots shown in \fig{fig:08} we conclude that $\theta_{Gauss}=D_f=2$ (the 2nd order phase transition) for a uniform hopping amplitudes ($\alpha=0$), and $\theta_{KPZ}=D_f=3$ (the 3rd order phase transition) for a descending hopping amplitudes with $\alpha=1/2$.

\end{appendix}

\newpage

\bibliography{ref.bib}

\end{document}